\documentstyle[aps,eqsecnum,epsf,rotate,prd,preprint,tighten]{revtex}

\newcommand{\ebox}[2]{\epsfxsize=#1 \epsfbox[10 30 560 590]{#2}}
\newcommand{\rf}[4]{{\em {#1}} {\bf #2}, #3 (#4)}

\newcommand{\pr}{Phys.\ Rev.\ }
\newcommand{\zf}{Z.\ Phys.\ }
\newcommand{\np}{Nucl.\ Phys.\ }
\newcommand{\beq}{\begin{equation}}
\newcommand{\eeq}{\end{equation}}
\newcommand{\beqa}{\begin{eqnarray}}
\newcommand{\eeqa}{\end{eqnarray}}
\newcommand{\half}{\frac{1}{2}}
\newcommand{\bra}{\langle}
\newcommand{\ket}{\rangle}
\newcommand{\Tr}{{\rm Tr}\,}
\renewcommand{\Re}{{\rm Re}\,}
\newcommand{\muhat}{\hat{\mu}}
\newcommand{\qhat}{\hat{q}}
\newcommand{\sss}{\scriptscriptstyle}
\newcommand{\Duv}{D_{\sss\rm UV}}
\newcommand{\Dir}{D_{\sss\rm IR}}
\newcommand{\csqdf}{\chi^2/N_{df}}
\newcommand{\err}[2]{\mbox{$\stackrel{\scriptstyle +#1}{\scriptstyle -#2}$}}

\begin{document}

\title{Asymptotic Scaling and Infrared Behavior of the Gluon Propagator}

\author{Derek B.\ Leinweber, 
Jon Ivar Skullerud\footnote{UKQCD Collaboration} and Anthony G.\ Williams}
\address{Special Research Centre for the Subatomic Structure of Matter
and Department of Physics and Mathematical Physics, University of
Adelaide, Adelaide SA 5005, Australia}

\author{Claudio Parrinello\footnotemark[1]}
\address{Department of Mathematical Sciences, University of Liverpool,
Liverpool L69 3BX, England}

\maketitle

\begin{abstract} 

The Landau gauge gluon propagator for the pure gauge theory is
evaluated on a $32^3\times 64$ lattice with a physical volume of
$(3.35^3\times 6.7)\ {\rm fm}^4$.  Comparison with two smaller lattices
at different lattice spacings allows an assessment of finite volume
and finite lattice spacing errors.  Cuts on the data are imposed to
minimize these errors.  Scaling of the gluon propagator is
verified between $\beta=6.0$ and $\beta=6.2$.  The tensor
structure is evaluated and found to be in good agreement with the
Landau gauge form, except at very small momentum values, where some
small finite volume errors persist.  A number of functional forms for
the momentum dependence of the propagator are investigated.  The form
$D(q^2)=\Dir+\Duv$, where $\Dir(q^2)\propto(q^2+M^2)^{-\eta}$ and $\Duv$
is an infrared regulated one-loop asymptotic form, is found to provide
an adequate description of the data over the entire momentum region
studied --- thereby bridging the gap between the infrared confinement
region and the ultraviolet asymptotic region.  The best estimate for
the exponent $\eta$ is $3.2\err{0.1}{0.2}\err{0.2}{0.3}$, where the
first set of errors represents the uncertainty associated with varying
the fitting range, while the second set of errors reflects the
variation arising from different choices of infrared regulator in
$\Duv$.  Fixing the form of $\Duv$, we find that 
the mass parameter $M$ is $(1020\pm 100)$ MeV.

\end{abstract}


\section{Introduction}
\label{sec:intro}

Over the years, the infrared behavior of the gluon propagator has
been studied using a variety of approaches, and with widely differing
results.  Gribov \cite{gribov} argued that by restricting the
functional integral to eliminate gauge copies, one would obtain a
gluon propagator which vanishes in the infrared.  Stingl \cite{stingl}
found that this solution was consistent with the gluon
Dyson--Schwinger equation (DSE), when ignoring the 4-gluon vertex and
placing certain restrictions on the remaining vertices.  Recent
studies of the coupled ghost and gluon DSEs \cite{hauck,bloch} support
this conclusion, in principle if not in detail.  On the other hand,
DSE studies of the gluon self-energy \cite{mandelstam,bbz,bp}
(ignoring the role of ghosts) have resulted in a gluon propagator
which is strongly enhanced in the infrared.  Occupying the `middle
ground' between these positions, Cornwall \cite{cornwall} has used a
gauge invariant `pinch technique' DSE to obtain a dynamical gluon
mass.  For a recent review of DSEs, see Ref.~\cite{cdr-agw}.

The infrared behavior of the gluon propagator is often considered to
be crucial to confinement.  Both the infrared-vanishing and the
infrared-enhanced solutions have been argued to provide mechanisms for
confinement.  It has even been argued \cite{hmr} that an
infrared-enhanced gluon propagator is a {\em necessary} condition for
confinement.  Clearly then, a settlement of this issue should allow
us to shed some light on the problem of confinement.

Lattice field theory provides a model-independent, {\em ab initio}
approach to QCD, and can in principle answer this question.
However, previous lattice studies of the gluon propagator in Landau gauge
\cite{bps,mms} have been inconclusive.  The reason for this is that
the lowest non-trivial momentum value accessible on a finite lattice
is inversely proportional to the length of the box.  The region of
interest is likely to be below 1 GeV.  Ref.~\cite{mms} used a lattice
with a spatial length of 2.5 fm and a length of 5 fm in the time
direction, giving access in principle to momentum values down to 250
MeV.  However, finite volume effects could be shown to be significant
at least up to approximately 500 MeV on this lattice, thereby casting doubt on
the validity of the results in the infrared.  In this study we
increase the lattice size to 3.35 fm in the spatial directions and 6.7
fm in the time direction, giving access to momenta deeper in the
infrared and significantly reducing finite volume effects.
Preliminary results can be found in Ref.~\cite{prev}.  We have also
compared the results for this lattice to those obtained on a smaller
volume and used anisotropies in the data to assess finite volume
effects.  However, an extrapolation to infinite volume has not been
attempted. 

The structure of this paper is as follows:  In Section
\ref{sec:gluon-lat} we present our method for calculating the gluon
propagator on the lattice, as well as the notation we use.  The
details of our simulations are given in Section \ref{sec:sim-params}.
In Section \ref{sec:cuts} we discuss how to handle finite volume and
finite lattice spacing artefacts.  The majority of our results can be
found in Section \ref{sec:results}.  Section \ref{sec:tensor}
discusses the tensor structure; in Section \ref{sec:asymptotics} the
asymptotic behavior is studied; and in Section \ref{sec:models} we
fit the gluon propagator as a function of momentum to
various functional forms.  Finally, in Section \ref{sec:discuss} we
discuss the significance of our results.

\newpage
\section{The gluon propagator on the lattice}
\label{sec:gluon-lat}

\subsection{Definitions and notation}
\label{sec:define}

The gauge links $U_\mu(x) \in $ SU(3) may be expressed in terms of the
continuum gluon fields as
\beq
U_\mu(x) = {\cal P} e^{ig_0\int_x^{x+\muhat}A_\mu(z)dz} 
= e^{ig_0 aA_\mu(x+\muhat/2)} + {\cal O}(a^3)\, .
\eeq
where ${\cal P}$ denotes path ordering.  From this, the dimensionless
lattice gluon field $A^L_{\mu}(x)$ may be obtained via
\beq
A^L_\mu(x+\muhat/2) = \frac{1}{2ig_0}\left(U_\mu(x)-U^{\dagger}_\mu(x)\right)
 - \frac{1}{6ig_0}\Tr\left(U_\mu(x)-U^{\dagger}_\mu(x)\right) \, ,
\label{eq:gluon-def}
\eeq
which is accurate to ${\cal O}(a^2)$. The discrete momenta $\qhat$
available on a finite, periodic volume of length $L_\mu$ in the $\mu$
direction, are given by
\beq
\qhat_\mu  = 
\frac{2 \pi n_\mu}{a L_\mu}, \qquad
n_\mu=0,\ldots,L_\mu-1 \, .
\eeq
The momentum space gluon field is
\beqa
A_\mu(\qhat) & \equiv & \sum_x e^{-i\qhat\cdot(x+\muhat/2)}
 A^L_\mu(x+\muhat/2) \nonumber \\
 & = & \frac{e^{-i\qhat_{\mu}a/2}}{2ig_0}\left[\left(U_\mu(\qhat)-U^{\dagger}_\mu(-\qhat)\right)
 - \frac{1}{3}\Tr\left(U_\mu(\qhat)-U^{\dagger}_\mu(-\qhat)\right)\right] , 
\eeqa
where $U_\mu(\qhat)\equiv\sum_x e^{-i\qhat x}U_\mu(x)$,
$A_\mu(\qhat)\equiv t^a A_{\mu}^a(\qhat)$, and $t^a$ are the
generators of the SU(3) Lie algebra.  This definition differs by a
term of ${\cal O}(a)$ from the one usually found in the literature,
where $U_\mu(x) = \exp(ig_0{A'}_\mu(x))$, which gives
${A'}_\mu(\qhat) = \exp(i\qhat_{\mu}a/2)A_\mu(\qhat) = A_\mu(\qhat) +
{\cal O}(a)$.  The dimensionless lattice gluon propagator
$D_{\mu\nu}^{L,ab}(\qhat)$ is defined by
\beq
\bra A_\mu^a(\qhat)A_\nu^b(-\qhat')\ket = V\delta(\qhat-\qhat')
D^{L,ab}_{\mu\nu}(\qhat) \, ,
\eeq
where $V$ is the lattice volume.

The continuum, infinite-volume gluon propagator in a covariant gauge
with gauge parameter $\xi$ has the form
\beq
D_{\mu\nu}^{ab}(q) =
(\delta_{\mu\nu}-\frac{q_{\mu}q_{\nu}}{q^2})\delta^{ab}D(q^2) 
 + \xi\frac{q_{\mu}q_{\nu}}{q^2}\delta^{ab}\frac{1}{q^2}
\, .
\label{eq:covgauge-prop}
\eeq
The scalar function $D(q^2)$ can be extracted from
$D_{\mu\nu}^{ab}(q)$ by
\beq
D(q^2) = \frac{1}{3}\left(\left[
\sum_{\mu}\frac{1}{8}\sum_{a}D_{\mu\mu}^{aa}(q)\right]
- \frac{\xi}{q^2}\right)\, .
\label{eq:scalar-prop}
\eeq
This expression is also valid on a finite volume, provided $q$ is not too
close to zero.  The finite volume induces an effective `mass'
$m\sim 1/L$ which becomes significant for $q$ sufficiently close to 0.
In this case, the most general form possible for the tensor structure is
\beq
D_{\mu\nu}^{ab}(q) =
\left(\delta_{\mu\nu}-\frac{h_{\mu\nu}(q)}{f(q^2)}\right)\delta^{ab}D(q^2) 
 + \xi\delta^{ab}\frac{h'_{\mu\nu}(q)}{g(q^2)}
\, ,
\label{eq:covgauge-finite-V}
\eeq
where, $f(q^2)\to q^2$, $g(q^2)\to q^4$, and $h_{\mu\nu}$ and  $h'_{\mu\nu}\to
q_{\mu}q_{\nu}$ for sufficiently large $q$, but $f(q^2)$ and $g(q^2)$
go to finite values for $q=0$.  In the following, we will work in the
Landau gauge, $\xi=0$, and we will only attempt fits to lattice data
for which finite size effects can be shown to be small.

A well-known lattice artefact is that the tree level propagator of a
massless scalar boson field does not reproduce the expected continuum
result of
\beq
D^{(0)}(q^2) = \frac{1}{q^2} ,
\label{eq:tree}
\eeq
but rather produces
\beq
D^{(0)}(\qhat) = \frac{1}{\sum_{\mu}(\frac{2}{a}\sin \qhat_\mu a/2)^2} \, .
\label{eq:lat-tree}
\eeq
Since QCD is asymptotically free, we expect that
$q^2 D(q^2)\to 1$ up to logarithmic
corrections as $q^2\to\infty$.
To ensure this result we work with a momentum variable
defined as\footnote{The momenta $q$ and $\qhat$ are often
defined the other way around in the lattice literature.  However, we feel it
is more instructive here to define $q$ as above, such that
lattice results reproduce the
continuum formula (\protect\ref{eq:covgauge-prop}) and the tree level
formula (\protect\ref{eq:tree}).}
\beq
q_{\mu} \equiv \frac{2}{a} \sin\frac{\qhat_{\mu} a}{2}\, .
\label{eq:latt-momenta}
\eeq
In the infrared region of greatest interest, the choice of $q$ vs.\
$\qhat$ makes little difference in the results.

\subsection{Renormalization}
\label{sec:renormalise}

The bare, dimensionless lattice gluon propagator $D^L(qa)$ is related
to the renormalized continuum propagator $D_R(q;\mu)$ via
\beq
a^2 D^L(qa) = Z_3(\mu,a) D_R(q;\mu) \, .
\label{eq:renorm-def}
\eeq
The renormalization constant $Z_3(\mu,a)$ is determined by imposing a
renormalization condition at some chosen renormalization scale $\mu$,
eg.,
\beq
D_R(q)|_{q^2=\mu^2} = \frac{1}{\mu^2} \, .
\label{eq:renorm-mom}
\eeq
The renormalized gluon propagator can be computed both
non-perturbatively on the lattice and perturbatively in the continuum
for choices of the renormalization point in the ultraviolet.  It can
then be related to the propagator in other continuum renormalization
schemes, such as $\overline{\rm MS}$.

\subsection{Gauge fixing}
\label{sec:gfix}

The lattice implementation of the Landau gauge is based on a
variational principle.  In continuum language, this can be seen by
defining for any generic field configuration $A_{\mu} (x)$ the
following functional on the group of gauge transformations:
\beq
F_{A}^{c} [g] = \parallel\!  A^g\!\parallel^2 =
\int d^4\!x \Tr (A^g_\mu(x))^2, 
\label{eq:gf-functional}
\eeq
where
\beq
A^g_\mu(x) = g^{-1}(x)A_\mu(x)g(x) - g^{-1}(x)\partial_\mu g(x) \,  \qquad 
g(x) \in {\rm SU(3)}.
\eeq
By considering gauge transformations of the form  
\beq
g'(x) = g(x) e^{i\omega(x)} = g(x) e^{it^a\omega^a(x)}
\eeq
and expanding to second order in $\omega$, it can be shown that
\beq
F^c_{A} [g'] = F^c_{A} [g] 
  - 2i \int d^4\!x \Tr(A^{g}_\mu(x)\partial_\mu\omega(x))
  - \int d^4\!x \Tr\sum_a\omega^a(x) (O[A^{g}]\omega)^a(x) \, + {\cal O} 
(\omega^3).
\label{eq:quad}
\eeq
This implies that $F_{A}^{c} [g]$ is stationary when $A^g_\mu (x)$
satisfies the Landau gauge condition $\partial \cdot A^g = 0$.  If
$A^g$ is in the Landau gauge, the operator appearing in the quadratic
term of Eq.~(\ref{eq:quad}) is $O[A^{g}] \equiv FP[A^{g}]$, i.e. the
Faddeev--Popov operator in the Landau gauge:

\beq
(FP [A^g] )^{ab}_{x y} = - \left( \partial \cdot \partial \ {\delta}^{ab} +
{f}^{abc} {A^g}_{\mu}^{c} (x) {\partial}_{\mu} \right) {\delta}^{4} (x - y).
\label{eq:FP}
\eeq 

Since configurations corresponding to local minima of $F_{A}^{c} [g]$
satisfy the gauge condition, Landau gauge fields may be constructed
from a generic configuration $A_{\mu}(x)$ by minimizing $F_{A}^{c}
[g]$.  This can be implemented in a quite straighforward way on the
lattice: a suitable discretization of $F_{A}^{c} [g]$ is given by
\beq
F_U^L [g] = 1-\sum_{\mu,x}\Re\Tr U^g_\mu(x) \, ,
\label{eq:lat-gf-functional}
\eeq
where
\beq
U^g_\mu(x) = g(x)U_\mu(x)g^{\dag}(x+\muhat).
\eeq 
$F_U^L [g]$ can be minimized numerically 
(for a review, see Ref.~\cite{gf-alg}), and the resulting link
configurations satisfy
\beq
\theta = \frac{1}{VN_c}\sum_x\theta(x) 
= \frac{1}{VN_c}\sum_x\Tr(\Delta(x)\Delta^\dagger(x)) = 0
\eeq
where $\Delta(x)$ is the lattice four-divergence of the gluon field,
\beq
\Delta(x) = \sum_\mu \left(A^L_\mu(x+\muhat/2)-A^L_\mu(x-\muhat/2)\right)
 \equiv \sum_\mu \left({A'}_\mu(x) - {A'}_\mu(x-\muhat)\right) \, .
\eeq

In momentum space, the lattice Landau gauge condition $\Delta(x)=0$
reads
\beq
\sum_\mu q_\mu A_\mu(q) = 0 \, ,
\label{eq:landau-condition}
\eeq
using the definition of $q$ in Eq.~(\ref{eq:latt-momenta}).  It is
worth noting that Eq.~(\ref{eq:landau-condition}) only holds if one
defines the gluon field according to Eq.~(\ref{eq:gluon-def}).  If the
asymmetric definition $A'$ is used instead, then
Eq.~(\ref{eq:landau-condition}) is replaced by
\beq
\sum_\mu \left(i\sin\qhat_\mu+1-\cos\qhat_\mu\right) {A'}_\mu(\qhat)
= 0 \, .
\eeq
In the limit $a \rightarrow 0$ the continuum Landau gauge condition is
recovered with ${\cal O} (a^2)$ corrections if one uses the field
defined in Eq.~(\ref{eq:gluon-def}) and with ${\cal O} (a)$
corrections if $A'$ is used.  This makes Eq.~(\ref{eq:gluon-def}) the
preferred definition.

Coming back to the continuum formulation, it is well known that in
non-abelian gauge theories, given a typical (regularized) field
configuration $A_{\mu} (x)$, the functional $F_{A}^{c} [g]$ will in
general have multiple stationary points. These correspond to distinct
configurations (Gribov copies), related to each other by gauge
transformations, which all satisfy the Landau gauge condition. This is
a consequence of the fact that the Faddeev-Popov operator
(\ref{eq:FP}) is not positive definite. In particular, it can be shown
that multiple local minima can occur, so that local minimization of
$F_{A}^{c} [g]$ does not fix the gauge uniquely.
This feature of the theory is preserved on the lattice
\cite{gribov-lat}, as it turns out
that $F_U^L [g]$ can have multiple stationary points (lattice Gribov
copies), and in particular multiple local minima.

Some possible solutions to this problem have been suggested in the
literature, mainly aiming to identify the global minimum of the
gauge-fixing functional (see for example Ref.~\cite{hdf}). At present,
the problem is still open. However, from the point of view of the
quantum theory, the relevant issue is to quantify the numerical impact
of the residual gauge freedom on gauge-fixed correlation functions. In
the framework of a Monte Carlo simulation, one may look for the
signature of gauge uncertainty as a ``noise'' effect, in addition to
the purely statistical uncertainty.  Previous studies
\cite{pptv,cucchieri1} indicate that this effect is negligible for most
gauge dependent quantities including the gluon
propagator.\footnote{The infrared behavior of the ghost propagator may
be more sensitive to the removal of Gribov copies \cite{cucchieri1}.}
For the purpose of the present investigation we shall therefore assume
that for the gluon propagator, the numerical uncertainty associated
with Gribov copies effects provides only a small contribution to the
overall error bars.

In the continuum formulation of the abelian gauge theory there is no Gribov
problem, as the Faddeev-Popov operator reduces to
a positive definite, field-independent one.  However, it is
interesting to notice that abelian Gribov copies may appear on the
lattice \cite{deForcrand:1991ux}, due to the structure of the lattice
Faddeev-Popov operator.

\section{Simulation parameters and methods}
\label{sec:sim-params}

The
details of the simulations are given in Table \ref{tab:sim-params}.
In short,
we analyze three lattices, two at $\beta=6.0$ and one at $\beta=6.2$,
 and denote these
as the `large', `small' and `fine' lattices respectively.
The gauge configurations are generated using a combination of
the over-relaxation and Cabibbo--Marinari algorithms.
All three lattices are fixed to Landau gauge
using a Fourier accelerated steepest descent algorithm \cite{cthd}. 

\begin{figure}[htbp]
\begin{center}
\setlength{\unitlength}{0.9cm}
\setlength{\fboxsep}{0cm}
\begin{picture}(14,8)
\put(0,0){\begin{picture}(7,7)\put(-1.5,0){\ebox{8.0cm}{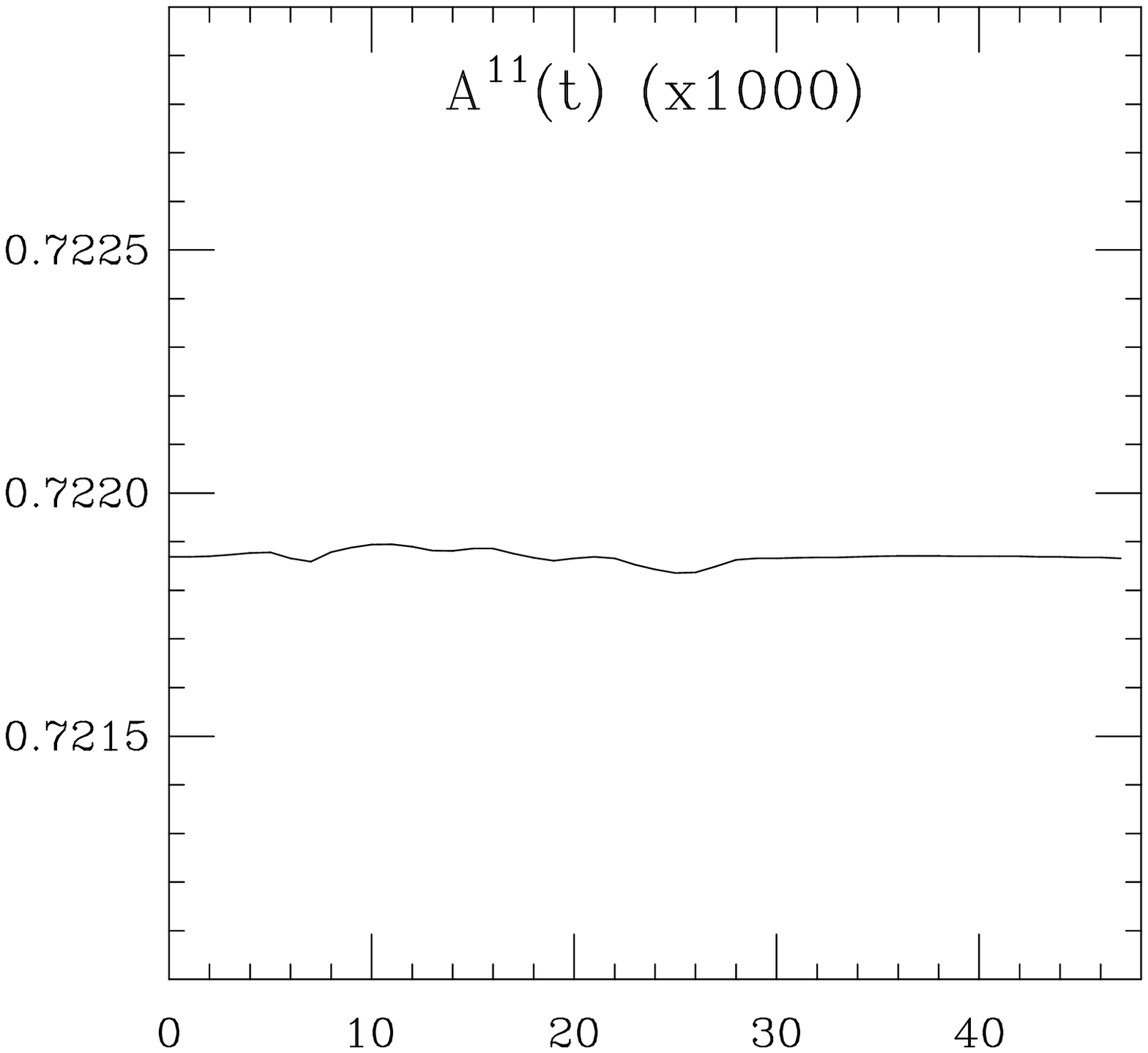}}\end{picture}}
\put(7,0){\begin{picture}(7,7)\put(0,0){\ebox{8.0cm}{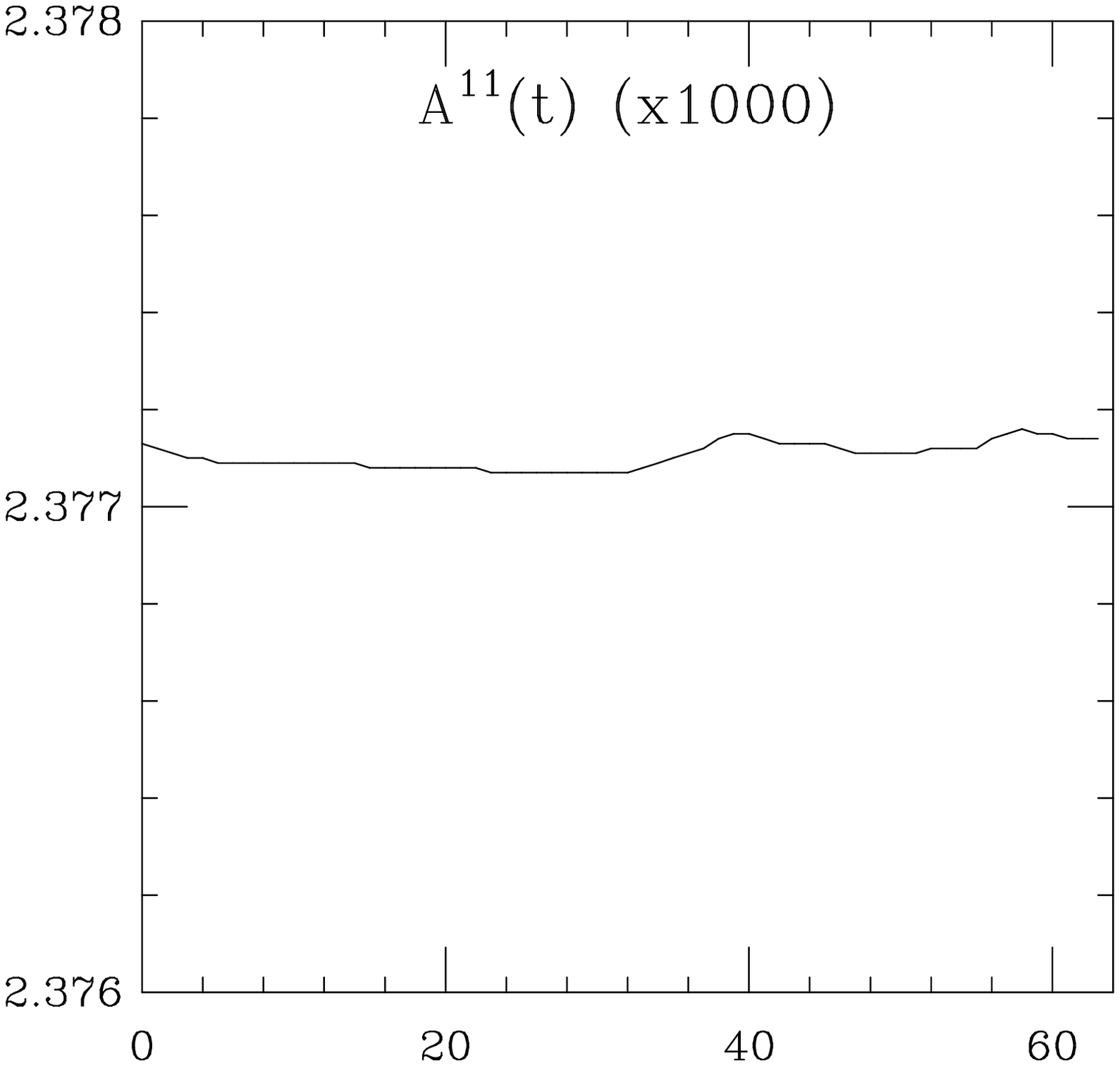}}\end{picture}}
\end{picture}
\end{center}
\caption{Plots of the (1,1) color component of
$\sum_{\vec{x}}A_4(\vec{x},t)$ as a function of $t$ for one gauge
fixed configuration on the small lattice (left), and on the large
lattice (right).}
\label{fig:a4}
\end{figure}

To double-check the gauge fixing we also consider
$\sum_{\vec{x}}A_4(\vec{x},t)$, which should be constant in time
when using periodic boundary conditions,
\beq
\partial_t \sum_{\vec{x}}A_4(\vec{x},t) = 
- \sum_{\vec{x}}\partial_i A_i(\vec{x},t) = 0 \, ,
\eeq
where we have introduced the shorthand notation
\beqa
\partial_t A_4(\vec{x}, t) & \equiv &  A^L_4(\vec{x},t+a/2) 
	- A^L_4(\vec{x},t-a/2) \\
\partial_i A_i(\vec{x}, t) & \equiv &  A^L_i(\vec{x} + \hat{e}_i/2,t) 
        - A^L_i(\vec{x} - \hat{e}_i/2,t)
\eeqa
In Fig.~\ref{fig:a4} we show typical values of 
$\sum_{\vec{x}}A_4(\vec{x},t)$ for both the
small and the large lattice.  As one can see, the time component of
the gluon field is constant to 1 part in 10000.  Note that the value
of one of the color components of the gluon field has no significance
in itself, although the fact that it is constant in time has.

\section{Finite size effects and anisotropies}
\label{sec:cuts}

\begin{figure}[p]
\begin{center}
\epsfysize=11.6truecm
\leavevmode
\rotate[l]{\vbox{\epsfbox{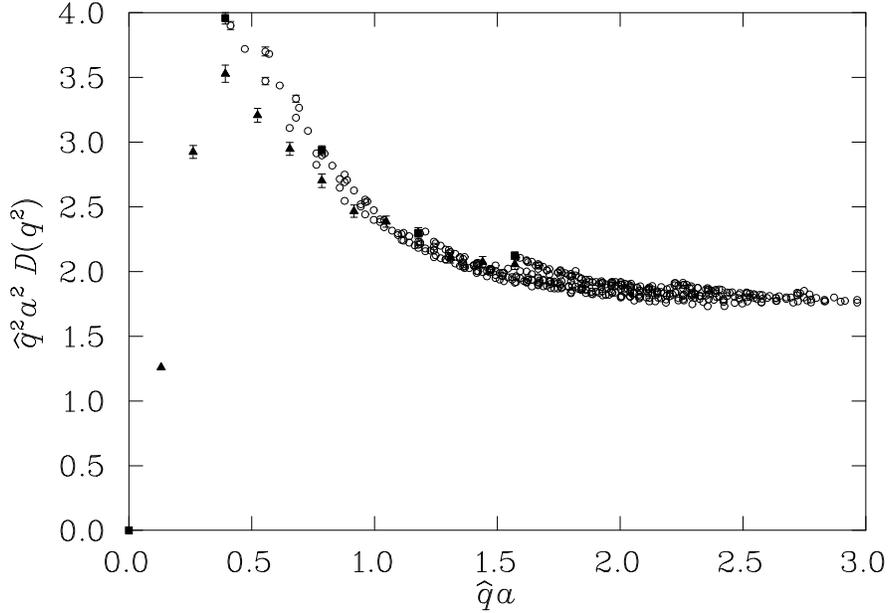}}}
\end{center}
\caption{The gluon propagator from the small lattice multiplied by
$\qhat^2 a^2$ plotted as a function of momenta $\qhat a$.  Values for each
momentum direction are plotted separately.  Only a $Z_3$ averaging has been
performed.  Filled squares denote momenta directed along spatial
axes, while filled triangles denote momenta directed
along the time axis.  Other momenta are indicated by open circles.}
\label{fig:alldata-small-qhat}
\end{figure}

We begin by considering the effect of the kinematic correction
introduced through the change of variables in Eq.~(\ref{eq:latt-momenta}).
In the absence of this correction, data in the high momentum region
are expected to exhibit significant anisotropy when shown as a
function of $\qhat$.  This is confirmed in
Fig.~\ref{fig:alldata-small-qhat}, which shows the gluon propagator
multiplied by $\qhat^2 a^2$ and plotted as a function of $\qhat a$.
Here and in the following, a $Z_3$ averaging is performed on the data,
where for example the momentum along (x,y,z,t) = (2,1,1,1) is averaged
with (1,1,2,1) and (1,2,1,1).

\begin{figure}[p]
\begin{center}
\epsfysize=11.6truecm
\leavevmode
\rotate[l]{\vbox{\epsfbox{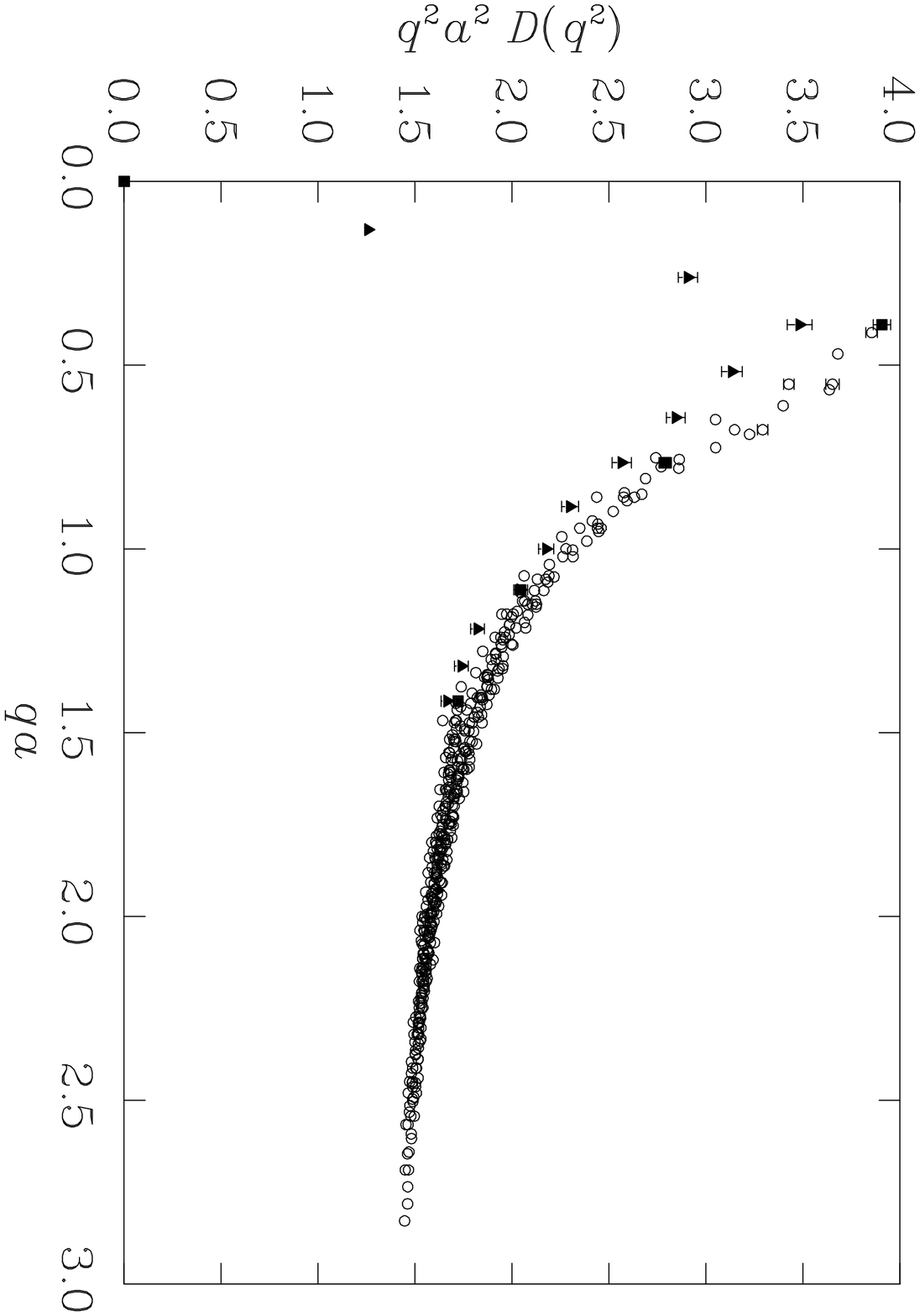}}}
\end{center}
\caption{The gluon propagator from the small lattice multiplied by
$q^2 a^2$ plotted as a function of momenta $qa$. The symbols are as in
Fig.~\protect\ref{fig:alldata-small-qhat}.}
\label{fig:alldata-small-q}
\end{figure}

\begin{figure}[p]
\begin{center}
\epsfysize=11.6truecm
\leavevmode
\rotate[l]{\vbox{\epsfbox{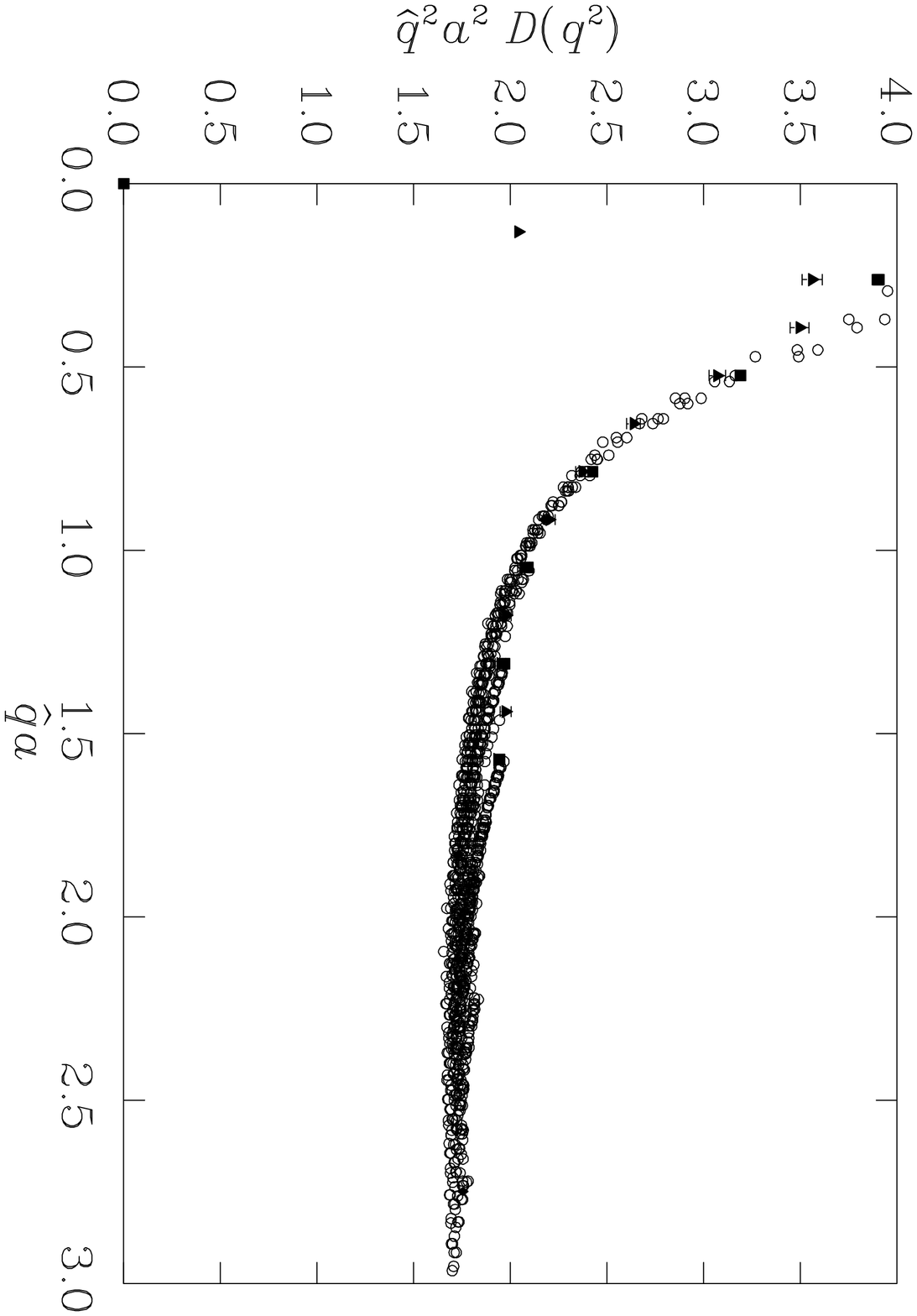}}}
\end{center}
\caption{The gluon propagator from the fine lattice multiplied by
$\qhat^2 a^2$ plotted as a function of momenta $\qhat a$.  The symbols
are as in Fig.~\protect\ref{fig:alldata-small-qhat}.}
\label{fig:alldata-fine-qhat}
\end{figure}

\begin{figure}[p]
\begin{center}
\epsfysize=11.6truecm
\leavevmode
\rotate[l]{\vbox{\epsfbox{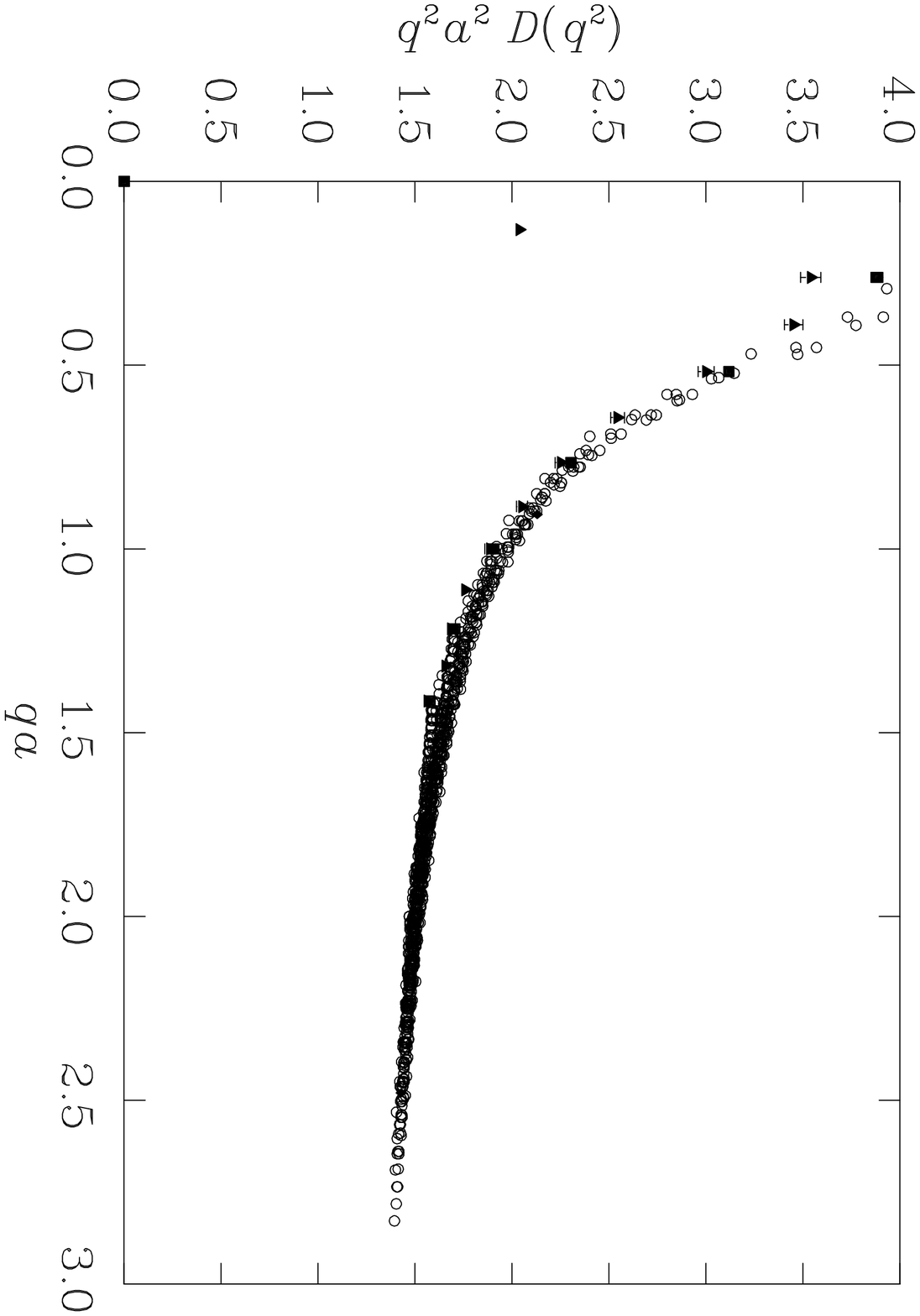}}}
\end{center}
\caption{The gluon propagator from the fine lattice multiplied by
$q^2 a^2$ plotted as a function of momenta $qa$.  The symbols are as in
Fig.~\protect\ref{fig:alldata-small-qhat}.}
\label{fig:alldata-fine-q}
\end{figure}

In Fig.~\ref{fig:alldata-small-q} the gluon propagator multiplied by
$q^2 a^2$ is displayed as a function of $qa$.  We see that the
kinematic correction results in a significant reduction in anisotropy
in the large momentum region, for $qa > 1.5$.  The effect of the
kinematic correction is even clearer for the fine lattice, as
displayed in Figs~\ref{fig:alldata-fine-qhat} and
\ref{fig:alldata-fine-q}.  We expect anisotropy arising from finite
lattice spacing artefacts to be reduced on this lattice, when the
lattice results are compared at the same physical value of $q$.
Rescaling these figures\footnote{Recall that the small to fine lattice
spacing ratio is $a_s/a_f = 1.4$.} and comparing them at the same physical
momenta shows a reduction in the anisotropy compared to the small
lattice in both cases.
However, this reduction is considerably smaller than the
one resulting from applying the kinematic correction on the fine
lattice.


At lower momenta, finite volume effects become significant.  These
effects are greatest when one or more of the momentum components is
zero.  Because of the unequal length of the time and spatial axes on
our lattices, there is a clear difference not only between on- and
off-axis points, but also between the points where
three of the components are zero, depending on whether or not one of
these lies along the `long' time axis.  In
figs.~\ref{fig:alldata-small-q} and \ref{fig:alldata-fine-q} this is
shown by the discrepancy between the filled squares (denoting momenta
along one of the spatial axes) and filled triangles (denoting momenta
along the time axis).

\begin{figure}[p]
\begin{center}
\epsfysize=11.6truecm
\leavevmode
\rotate[l]{\vbox{\epsfbox{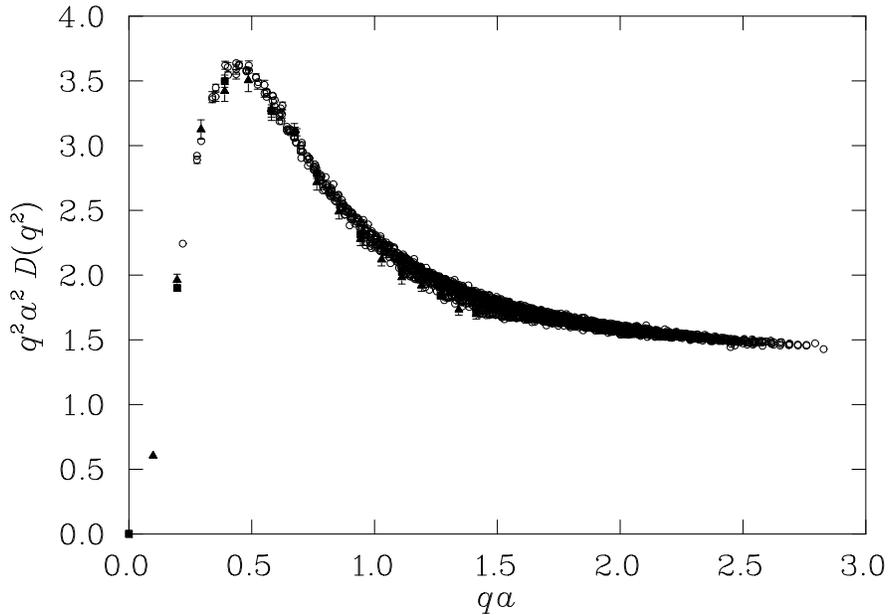}}}
\end{center}
\caption{The gluon propagator from the large lattice multiplied by
$q^2 a^2$ plotted as a function of momenta $qa$.  Values for each
momentum direction are plotted separately.  Only a $Z_3$ averaging has
been performed for the data shown in this figure.  Plotting symbols
are as in Fig.~\protect\ref{fig:alldata-small-qhat}.  Finite volume errors are
greatly reduced compared to the results from the smaller lattice, as
displayed by the overlap of points obtained from spatial and time-like
momentum vectors.  However, significant anisotropy is apparent for
larger momenta.}
\label{fig:alldata-large}
\end{figure}

Fig.~\ref{fig:alldata-large} displays the gluon propagator data for all
momentum directions and values on the large lattice, using the
kinematic correction.  Again, only a
$Z_3$ averaging has been performed.  Examination of the infrared
region indicates that finite volume artefacts are very small on the
large lattice.  In particular, the agreement between purely spatial
(filled squares) and time-like momentum vectors (filled triangles) at
$qa = 0.20$ appears to indicate that finite size effects are
relatively small here.


Some residual anisotropy remains for both the large and small lattices
at moderate momenta around $q a \sim 1.5$, despite including the
kinematic correction of Eq.~(\ref{eq:latt-momenta}).  This anisotropy is
clearly displayed in Fig.~\ref{fig:alldata-small-q} by the filled
squares and triangles denoting momenta directed along lattice axes
lying below the majority of points from off-axis momenta for $qa\sim
1.4$.  Since tree-level O(4) breaking effects should be
removed by the kinematic correction, 
the remaining anisotropy
appears to have its origin in quantum effects beyond tree level.
This anisotropy is significantly reduced for the fine lattice,
indicating that it is an effect of finite lattice spacing errors as
opposed to finite volume errors.  The fact that it occurs at the same
momentum values and with the same magnitude on both the large and
small lattices at $\beta=6.0$ lends further support to this
interpretation.

In order to minimize lattice artefacts for large momentum components
we select momentum vectors lying within a cylinder directed along the
diagonal $(x,y,z,t) = (1,1,1,1)$ of the lattice.  This allows one to
access the largest of momenta with the smallest of components.  We
calculate the distance $\Delta\qhat$ of the momentum vector $\qhat$
from the diagonal using
\beq
\Delta\qhat = |\qhat|\sin\theta_{\qhat} \, ,
\label{eq:deltaq-def}
\eeq
where the angle $\theta_{\qhat}$ is given by
\beq
\cos\theta_{\qhat} = \frac{\qhat\cdot\hat{n}}{|\qhat|} \, ,
\label{eq:angle-def}
\eeq
and $\hat{n} = \half(1,1,1,1)$ is the unit vector along the diagonal.

\begin{figure}[p]
\begin{center}
\epsfysize=11.6truecm
\leavevmode
\rotate[l]{\vbox{\epsfbox{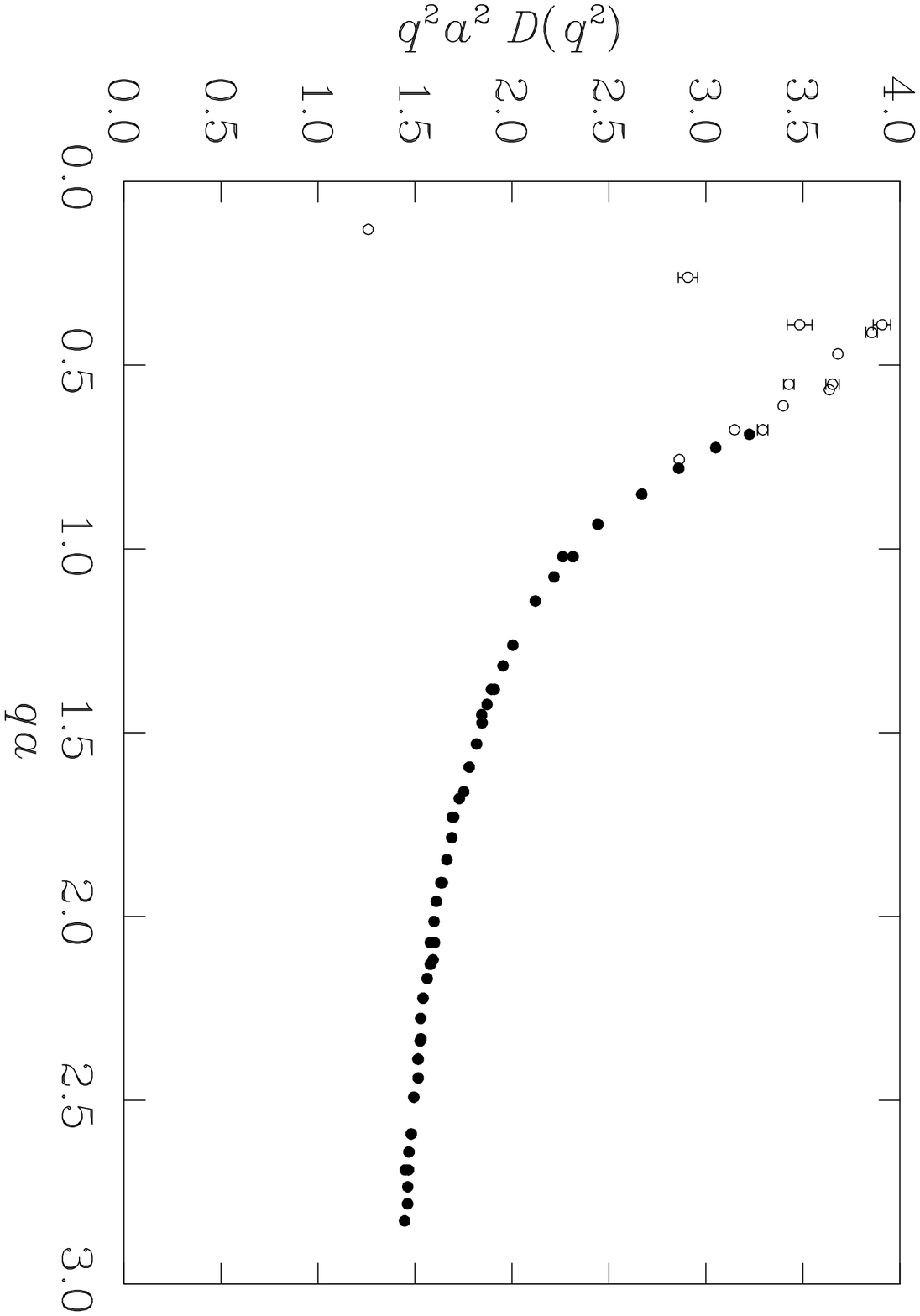}}}
\end{center}
\caption{The gluon propagator from the small lattice multiplied by
$q^2 a^2$.  The points displayed in this plot lie within a cylinder of
radius $\Delta\qhat a < 1\!\times\!2\pi/16$ directed along the
diagonal $(x,y,z,t) = (1,1,1,1)$ of the lattice.  The solid points
also lie within a cone of $20^\circ$ measured from the diagonal at the
origin.}
\label{fig:small-cuts}
\end{figure}

\begin{figure}[p]
\begin{center}
\epsfysize=11.6truecm
\leavevmode
\rotate{\vbox{\epsfbox{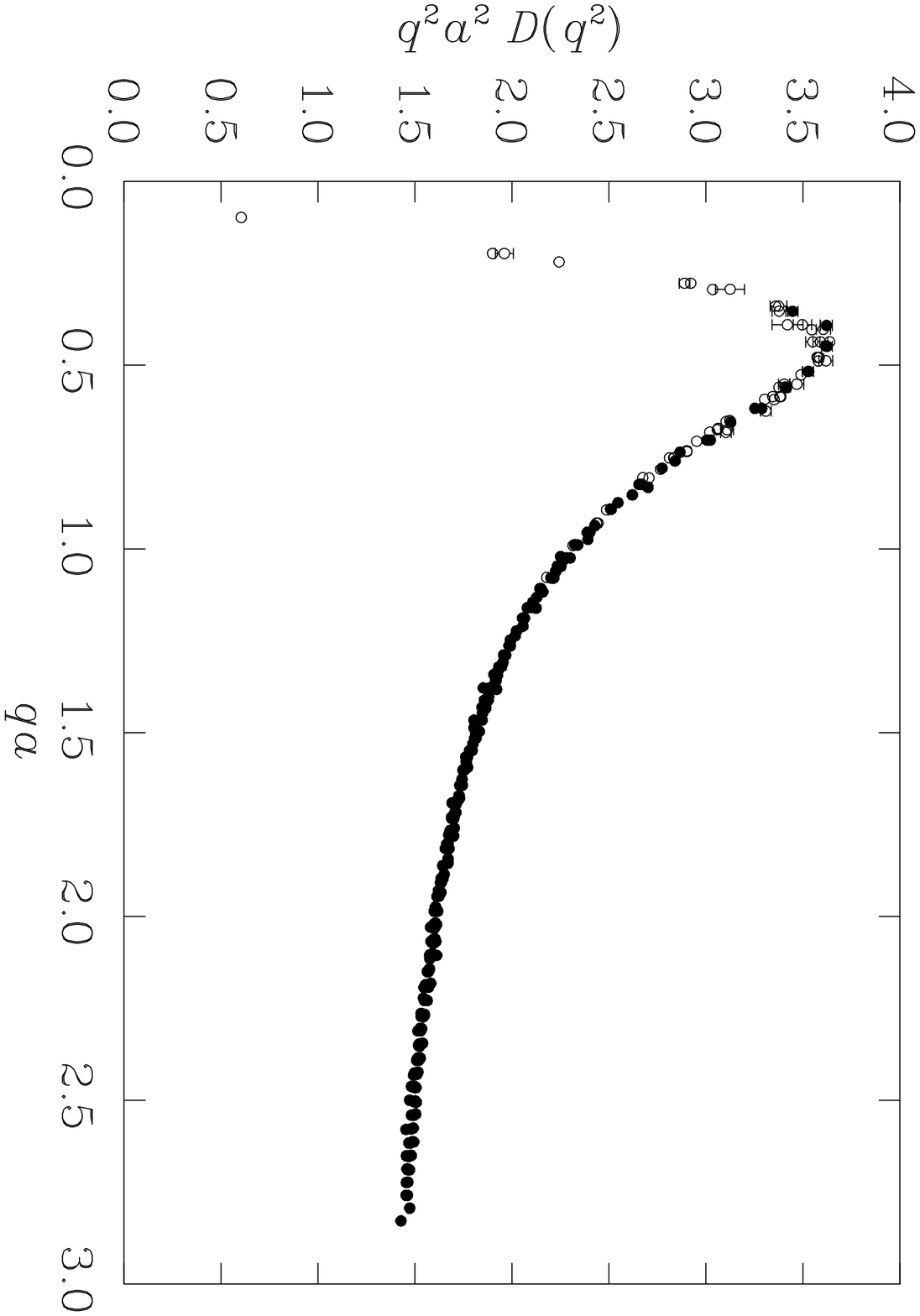}}}
\end{center}
\caption{The gluon propagator from the large lattice multiplied by
$q^2 a^2$.  The points displayed in this plot lie within a cylinder of
radius $\Delta\qhat a < 2\!\times\!2\pi/32$ directed along the
diagonal of the lattice.  The solid points also lie within a cone of
$20^\circ$ measured from the diagonal at the origin.}
\label{fig:large-cuts}
\end{figure}

\begin{figure}[p]
\begin{center}
\epsfysize=11.6truecm
\leavevmode
\rotate{\vbox{\epsfbox{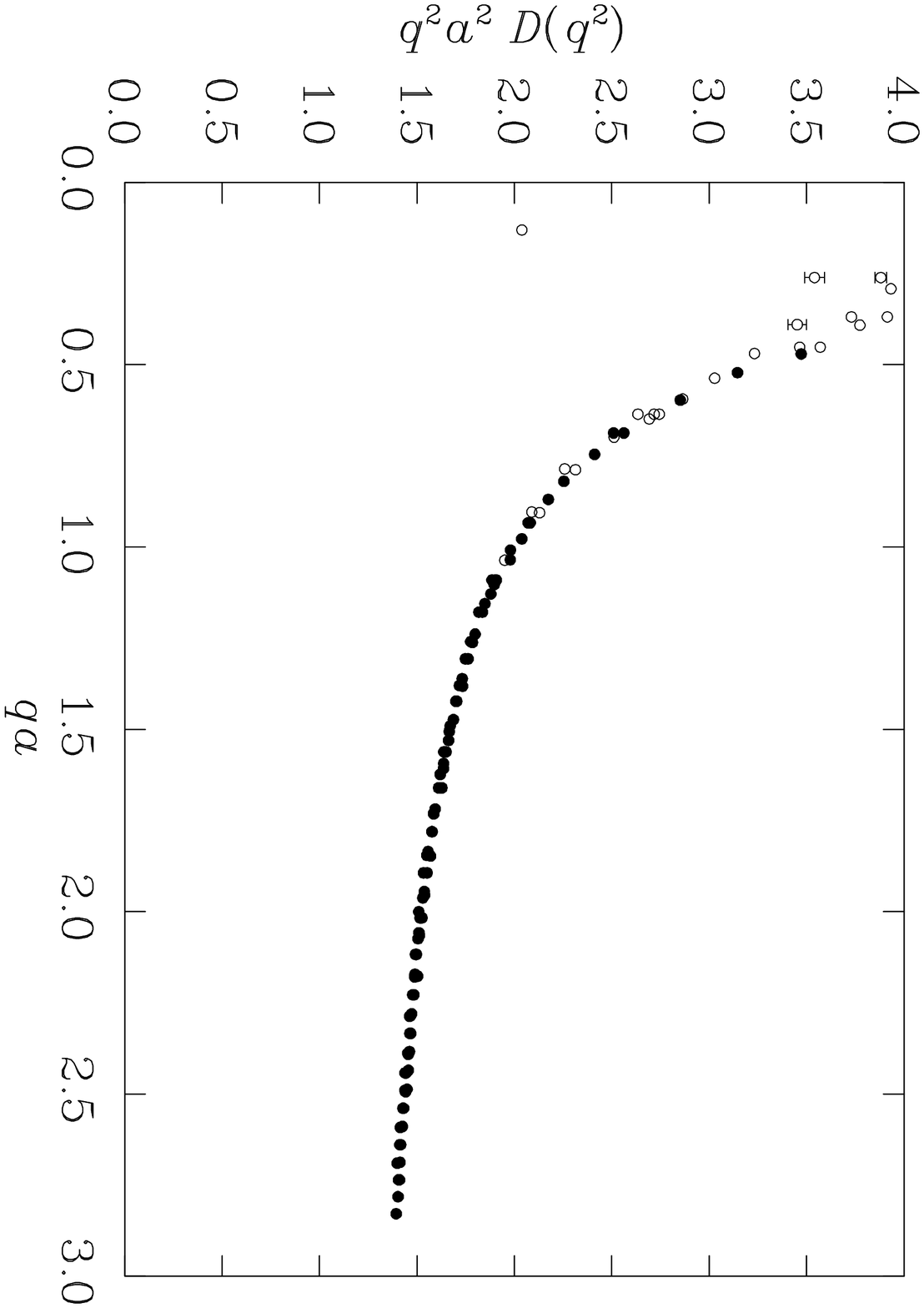}}}
\end{center}
\caption{The gluon propagator from the fine lattice multiplied by
$q^2 a^2$.  The points displayed in this plot lie within a cylinder of
radius $\Delta\qhat a < 1.5\!\times\!2\pi/32$ directed along the
diagonal of the lattice.  The solid points also lie within a cone of
$20^\circ$ measured from the diagonal at the origin.}
\label{fig:fine-cuts}
\end{figure}

On the
small lattice, we found the selection of a cylinder with a radius of
one spatial momentum unit ($\Delta\qhat a < 1\!\times\!2\pi/L_s$,
where $L_s$ is the number of sites along a spatial axis) provides a
reasonable number of points falling along a single curve for large
momenta.  The data surviving this cut are displayed in
Fig.~\ref{fig:small-cuts}.
For the large lattice the corresponding physical cut dictates that all momenta
must lie within a cylinder of radius two spatial momentum units
directed along the lattice diagonal.  Fig.~\ref{fig:large-cuts}
displays the data surviving this cut.  Fig.~\ref{fig:fine-cuts}
shows the data surviving the corresponding cut on the fine lattice,
using a radius of 1.5 momentum units, which provides a similar
physical radius.

This cut does not address the large finite volume errors
surviving in Fig.~\ref{fig:small-cuts}.  To remove these
problematic momenta, we consider a further cut designed to remove
momentum vectors which have one or more vanishing components.  This is
implemented by keeping only momentum directions that lie within a
certain angle $\theta_{max}$ from the diagonal, ie., by keeping 
$\theta_{\qhat} < \theta_{max}$ where $\theta_{\qhat}$ is given by
Eq.~(\ref{eq:angle-def}).  We found that a cone of semivertex angle
$\theta_{max} = 20^\circ$ was sufficient to
provide a set of points lying along a smooth curve.  The solid points
in Fig.~\ref{fig:small-cuts} represent these data. 


Since finite volume errors on the large lattice are small, it is not
necessary to impose the additional cone cut there.  However, it is 
interesting to
note that even with this conservative cut, illustrated by the solid
points in Fig.~\ref{fig:large-cuts}, the turnover in $q^2a^2 \,
D(q^2)$ in the infrared region is still observed.

\section{Results}
\label{sec:results}

\subsection{Tensor structure}
\label{sec:tensor}

Using the Landau gauge condition (\ref{eq:landau-condition}), we can
infer that the lattice gluon propagator $D^L_{\mu\nu}(q)\equiv
\frac{1}{8}\sum_a D^{L,aa}_{\mu\nu}(q)$ should have the following tensor
structure, mirroring the continuum form (\ref{eq:covgauge-prop}):
\beq
D^L_{\mu\nu}(q) =
(\delta_{\mu\nu}-\frac{q_{\mu}q_{\nu}}{q^2})D^L(q^2) \, .
\label{eq:landau-prop}
\eeq
By studying the tensor structure of the gluon propagator, we may be
able to determine how well the Landau gauge condition is satisfied,
and also discover violations of continuum rotational invariance.
The tensor structure may be evaluated directly by taking the ratios of
different components of $D^L_{\mu\nu}(q)$ for the same value of $q$.
The results for moderate to high momentum values, where we expect
Eq.~(\ref{eq:landau-prop}) to be valid, are summarized in Tables
\ref{tab:tensor-small}--\ref{tab:tensor-fine}, and compared to what
one would expect from Eq.~(\ref{eq:landau-prop}), and to what one would
obtain by replacing $q$ with $\qhat$ in Eq.~(\ref{eq:landau-prop}).  For
the small and fine lattices, we have also evaluated the tensor
structure using the unfavored asymmetric definition $A'$ of the gluon
field.

\label{tensortables}


The selected momentum values in Tables
\ref{tab:tensor-small}--\ref{tab:tensor-fine} are not an exhaustive
list, but are representative of the respective momentum regimes.  It
is clear from  these tables
that our numerical data are consistent with the expectation
from Eq.~(\ref{eq:landau-prop}).  In particular, where two of the
components of $q$ are zero, this relation is satisfied with a very
high degree of accuracy.  Where 3 or 4 of the components are non-zero,
the errors are larger, but in most cases smaller than 10\%.
We can also see that in general, the
asymmetric definition $A'$ of the gluon field gives results which are
inconsistent with this form.

At very low momentum values, we expect finite volume effects to lead
to violations of the infinite-volume continuum-limit form
(\ref{eq:landau-prop}).  Tables
\ref{tab:tensor-small-lowq}--\ref{tab:tensor-large-lowq} show selected
ratios of components for the lowest momentum values, displaying, in
some cases, significant violations of this form.  In particular, the
ratio of the $\mu=4$ (time) component of the diagonal
$D^L_{\mu\mu}(q)$ to the other diagonal components is considerably
larger than what one would get from Eq.~(\ref{eq:landau-prop}).  The
discrepancy is smaller on the large lattice than on the other two
lattices, but is still significant at these lowest momenta.  This
gives us a more rigorous test of finite volume effects than what we
could obtain by inspection in Section \ref{sec:cuts}, where finite
volume effects were not obvious for the large lattice.

At zero momentum Eq.~(\ref{eq:landau-prop}) is not well-defined, and the
finite volume replacement Eq.~(\ref{eq:covgauge-finite-V}) (with
$\xi=0$) must be used instead.  The exact behavior of the functions
$h_{\mu\nu}(q)$ and $f(q^2)$ with $q$ and $V$ is not known, but any
deviation from $D_{\mu\nu}(q=0) \propto \delta_{\mu\nu}$ must be due
to finite volume effects in $h_{\mu\nu}$ of Eq.~(\ref{eq:covgauge-finite-V}).  Table
\ref{tab:tensor-zeromom} shows the ratios of the diagonal elements for
our three lattices.  As we can see, the $\mu=4$ component is in all
cases much smaller than the other three components, although the
discrepancy is considerably reduced from the small to the large
lattice.  The small and the fine lattice have a ratio $D_{ii}/D_{44}$
of 3 and 2 respectively, which is equal to the ratio $L_t/L_i$.  For
the large lattice, with $L_t/L_i=2$, $D^L_{ii}/D^L_{44}\approx 1.4$
indicating the reduction of finite volume errors at zero momentum.


\subsection{Asymptotic behavior}
\label{sec:asymptotics}

The asymptotic behavior of the renormalized gluon propagator in the
continuum is given to one-loop level by \cite{mandelstam,da-z}
\beq
D_R(q^2;\mu) \equiv D_{\rm bare}(qa)/Z_3(\mu,a)
\sim \frac{Z}{q^2}\left(\half\ln(q^2/\Lambda^2)\right)^{-d_D} \, ,
\label{model:asymptotic}
\eeq
where the constant $Z$ depends on the renormalization scheme and the
renormalization point $\mu$, and\footnote{This expression differs by a
factor of 2 from the (incorrect) expression given in
Ref.~\cite{marc-pag}, which is also quoted in Ref.~\cite{cdr-agw}.}
\beq
d_D=\frac{39-9\xi-4N_f}{2(33-2N_f)} \, .
\eeq
In the case we are studying here, both the gauge parameter $\xi$ and
the number of fermion flavors $N_f$ are zero, so $d_D=13/22$.

\subsubsection{Fits to the asymptotic form}
\label{sec:asym-fits}

We have fitted the data, with the kinematic correction, for all our
three lattices to the asymptotic form in Eq.~(\ref{model:asymptotic})
for values of $q$ above $\sim 2.7$GeV.  For the large lattice, we have
used the data surviving the cylindrical cut, while for the other two
lattices, both the cylindrical and cone cuts have been imposed.  Table
\ref{tab:asymptotic_fits} shows the parameter values for the most
inclusive of those fits.  Other regimes are selected to
facilitate comparisons between the three lattices.  The largest region
providing $\csqdf \simeq 1$ is also indicated.


We see that the asymptotic form fits the data quite well, although the
relatively high $\chi^2$ for the fits beginning at $q\sim 2.7$ GeV may
be taken as a sign that there are still significant nonperturbative
and/or higher loop contributions to the propagator at this momentum scale.
The values for the scale parameter $\Lambda$ are reasonably consistent
for the two $\beta$-values, although the variation in $\Lambda$
between different lattices and fit ranges indicates that the one-loop
perturbative form is still not valid even at $q^2 = 25{\rm GeV}^2$.

\subsubsection{Matching results for the two lattice spacings}

Since the renormalized propagator $D_R(q;\mu)$ is independent of the
lattice spacing when the lattice spacing is fine enough (i.e., in the
scaling regime), we can use Eq.~(\ref{eq:renorm-def}) to derive a simple,
$q$-independent expression for the ratio of the unrenormalized
lattice gluon propagators at the same physical value of $q$:
\beq
\frac{D^L_f(qa_f)}{D^L_c(qa_c)} = 
\frac{Z_3(\mu,a_f)D_R(q;\mu)/a_f^2}{Z_3(\mu,a_c)D_R(q;\mu)/a_c^2}
= \frac{Z_f}{Z_c}\frac{a_c^2}{a_f^2}
\label{eq:gluon_match_ratio}
\eeq
where the subscript $f$ denotes the finer lattice ($\beta=6.2$ in this
study) and the subscript $c$ denotes the coarser lattice
($\beta=6.0$).  We can use this relation to study directly the scaling
properties of the lattice gluon propagator by matching the data for
the two values of $\beta$.  This matching can be performed by
adjusting the values for the ratios $R_Z = Z_f/Z_c$ and $R_a =
a_f/a_c$ until the two sets of data lie on the same curve.  
It should be emphasized that this procedure matches the
lattice data directly, and does not depend on a
functional form of the gluon propagator.

In this study, we have used the fine and small lattices 
to perform this matching, as they have similar physical volumes.  The
combination of cylindrical and cone cut has been applied to both data
sets. 
We have implemented the matching by making a linear interpolation of the
logarithm of the data plotted against the logarithm of the momentum
for both data sets.  In this way the scaling of the momentum is
accounted for by shifting the fine lattice data to the right by an
amount $\Delta_a$ as follows
\beq
\ln D^L_c( \ln(qa_c) ) = \ln D^L_f( \ln(qa_c) - \Delta_a ) + \Delta_Z
\label{eq:match-data}
\eeq
Here $\Delta_Z$ is the amount by which the fine lattice data must be
shifted up to provide the optimal overlap between the two data sets.
The matching of the two data sets has been performed for values of
$\Delta_a$ separated by a step size of 0.001.  $\Delta_Z$ is
determined for each value of $\Delta_a$ considered, and the optimal
combination of shifts is identified by searching for the global
minimum of $\csqdf$.  The ratios $R_a$ and $R_Z$ are related
to $\Delta_a$ and $\Delta_Z$ by
\beq
R_a = e^{-\Delta_a}\, , \hspace{2.0em} R_Z = R_a^2 e^{-\Delta_Z} \, .
\eeq

\begin{figure}[p]
\begin{center}
\epsfysize=11.6truecm
\leavevmode
\rotate[l]{\vbox{\epsfbox{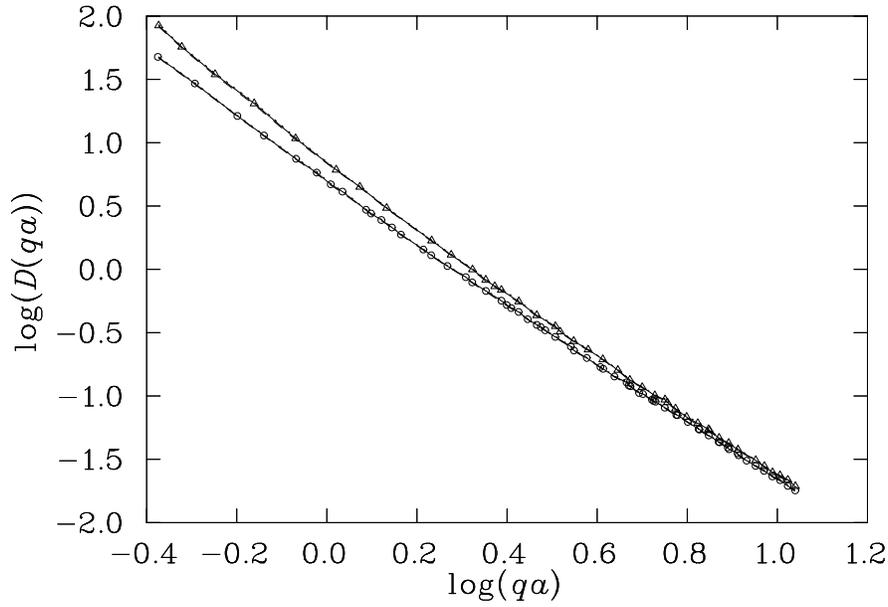}}}
\end{center}
\caption{The dimensionless, unrenormalized gluon propagator as a
function of $\ln(qa)$ for the two values of $\beta$.  The triangles
denote the data for the small (coarse) lattice at $\beta=6.0$, while the
circles denote the data for $\beta=6.2$.  The lines represent linear
interpolations between the data points.}
\label{fig:compare_data_lattmom}
\end{figure}
\begin{figure}[p]
\begin{center}
\epsfysize=11.6truecm
\leavevmode
\rotate[l]{\vbox{\epsfbox{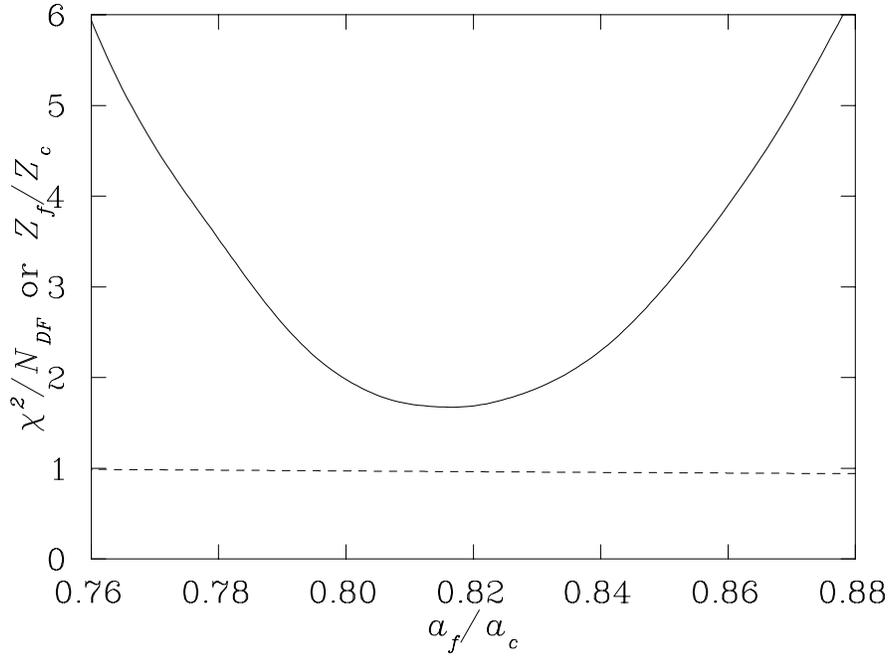}}}
\end{center}
\caption{$\chi^2$ per degree of freedom as a function of the ratio of
lattice spacings for matching the small and fine lattice data, using 
$\qhat$ as the momentum variable.  The dashed line indicates the
ratio $R_Z$ of the renormalization constants.}
\label{fig:match_spacing_contmom}
\end{figure}

Figure~\ref{fig:compare_data_lattmom} shows the data for both lattice
spacings as a function of $qa$ before shifting.  In
Fig.~\ref{fig:match_spacing_contmom} we present the result of the
matching using $\qhat$ as the momentum variable.  The minimum value
for $\csqdf$ of about 1.7 is obtained for $R_a\sim 0.815$.  This
value for $R_a$ is considerably higher than the value of $0.716\pm
0.040$ obtained from an analysis of the static quark potential in
Ref.~\cite{bs}.  From this discrepancy, as well as the relatively high
value for $\csqdf$, we may conclude that the gluon propagator,
taken as a function of $\qhat$, does not exhibit scaling behavior for
the values of $\beta$ considered here.

\begin{figure}[p]
\begin{center}
\epsfysize=11.6truecm
\leavevmode
\rotate[l]{\vbox{\epsfbox{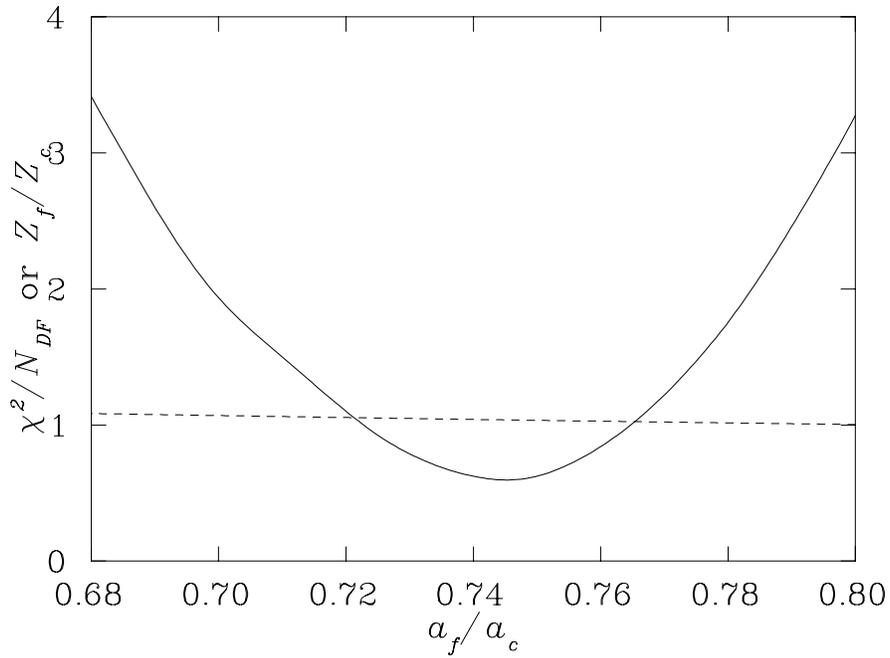}}}
\end{center}
\caption{$\chi^2$ per degree of freedom as a function of the ratio of
lattice spacings for matching the small and fine lattice data, using 
$q$ as the momentum variable.}
\label{fig:match_spacing_lattmom}
\end{figure}

Fig.~\ref{fig:match_spacing_lattmom} shows the result of the matching
using $q$ as the momentum variable.  We can see immediately that this
gives much more satisfactory values both for $\csqdf$ and for
$R_a$.  The minimum value for $\csqdf$ of 0.6 is obtained for
$R_a=0.745$.  Taking a confidence interval where $\csqdf <
\chi^2_{\rm min} + 1$ gives us an estimate of $R_a=0.745\err{32}{37}$,
where the errors denote the uncertainty in the last digits.
This is fully compatible with the value of $0.716\pm 0.040$ obtained from
Ref.~\cite{bs}.  The corresponding estimate for the ratio of the
renormalization constants is $R_Z = 1.038\err{26}{21}$.  That $R_Z
\geq 1$ is consistent with what one would expect from
continuum perturbation theory.


\subsection{Model functions}
\label{sec:models}

Having verified scaling in our lattice data over the entire range
of $q^2$ considered, we will now proceed with model fits.
We have considered a number of functional forms, based on a variety of
theoretical
suggestions from the literature.  All these forms, as well as the new
models we have constructed in this study, include an overall dimensionless
renormalization parameter $Z$.  This parameter is not equal to the
renormalization constant $Z_3$, although the two can be related for
each individual model.

We introduce an infrared-regulated version $L(q^2,M)$ of
the one-loop logarithmic correction given by
Eq.~(\ref{model:asymptotic}) in order to ensure that these models have
the correct leading ultraviolet behavior.  This is given by
\beq
L(q^2,M) \equiv 
\left[\half\ln\left((q^2+M^2)(q^{-2}+M^{-2})\right)\right]^{-d_D} \, .
\label{eq:uvlog-def}
\eeq
The factor $q^{-2}+M^{-2}$ ensures that $L(q^2,M)$ is properly
regulated in the infrared.

For simplicity of presentation of the models, all model formulae are
to be understood as functions of dimensionless quantities (scaled by
the appropriate powers of the lattice spacing $a$).  The models
considered here are:

{\bf Gribov \cite{gribov}}
\beq
D^L(q^2) = \frac{Zq^2}{q^4+M^4}L(q^2,M)
\label{model:lita}
\label{model:gribov}
\eeq

{\bf Stingl \cite{stingl}}
\beq
D^L(q^2) = \frac{Zq^2}{q^4+2A^2q^2+M^4}L(q^2,M)
\label{model:stingl}
\eeq

{\bf Marenzoni \cite{mms}}
\beq
D^L(q^2) = \frac{Z}{(q^2)^{1+\alpha}+M^2}
\label{model:marenzoni}
\label{model-first}
\eeq

{\bf Cornwall I \cite{cornwall}}
\beq
D^L(q^2) = Z\left[(q^2+M^2(q^2))\ln\frac{q^2+4M^2(q^2)}{\Lambda^2}\right]^{-1}
\label{model:cornwall}
\eeq
where
\[
M(q^2) = M\left\{\frac{\ln\frac{q^2+4M^2}{\Lambda^2}}
{\ln\frac{4M^2}{\Lambda^2}}\right\}^{-6/11}
\]

{\bf Cornwall II \cite{cornwall-priv}}
\beq
D^L(q^2) = Z\left[(q^2+M^2))\ln\frac{q^2+4M^2}{\Lambda^2}\right]^{-1}
\label{model:cornwall1}
\eeq

{\bf Cornwall III \cite{cornwall-priv}}
\beq
D^L(q^2) = \frac{Z}{q^2+Aq^2\ln(q^2/M^2)+M^2}
\label{model20}
\label{model:cornwall2}
\label{model:litz}
\eeq

Here, $Z, M, A, \Lambda$ and $\alpha$ are parameters to be optimized
in the fit.  In addition, we study the following 4-parameter forms:

{\bf Model A:}
\beq
D^L(q^2) = Z\left[\frac{AM^{2\alpha}}{(q^2+M^2)^{1+\alpha}} + 
\frac{1}{q^2+M^2}L(q^2,M)\right]
\label{modelA}
\eeq

{\bf Model B:}
\beq
D^L(q^2) = Z\left[\frac{AM^{2\alpha}}{(q^2)^{1+\alpha}+(M^2)^{1+\alpha}} +
\frac{1}{q^2+M^2}L(q^2,M)\right]
\label{modelB}
\eeq

{\bf Model C:}
\beq
D^L(q^2) = Z\left[\frac{A}{M^2}e^{-(q^2/M^2)^{\alpha}} + 
\frac{1}{q^2+M^2}L(q^2,M)\right]
\label{modelC}
\label{model-last}
\eeq

We have also considered special cases of the three forms
(\ref{modelA})--(\ref{modelC}), with specific
values for the exponent $\alpha$.  All these models are constructed to
exhibit the asymptotic behavior of Eq.~(\ref{model:asymptotic}).

\subsection{Numerical results}
\label{sec:numbers}

The fits are performed to the large lattice data using the
cylindrical cut, and excluding the first point (at $qa\sim 0.1$),
which may be sensitive to the volume of the lattice.  To balance the
sensitivity of the fit over the available range of $qa$, we have
averaged adjacent lattice momenta lying within $\Delta qa < 0.005$.

In order to determine the stability of the fits, we have varied the
starting point and width of the fit.  After averaging over adjacent
momenta, the data points are numbered 1,2,\ldots,142.  The starting
point has been incremented in steps of 2, and for each starting point
the width has been varied in steps of 2 between the minimum possible
width (ie, the number of parameters plus 1) and the maximum width.
The statistical uncertainty in the parameters is determined using a
jackknife procedure \cite{efron}.  Since the number of points in most
of the fits is larger than the number of configurations, we have not
been able to compute $\chi^2$ using the full covariance matrix
\cite{cmi}.  However, in the cases where this is possible, the results
are compatible with those achieved using the `na\"{\i}ve' $\chi^2$.


Table~\ref{tab:chisq_litmodels} shows the values for $\csqdf$
for each of the models (\ref{model:lita})--(\ref{model:litz}).
Unfortunately, none of these models succeed in providing an acceptable
fit over the entire available momentum range.  In the case of
Marenzoni's model (\ref{model:marenzoni}), this is not surprising, since this
model does not have the correct asymptotic behavior.  Our models,
Model A and Model B, are constructed as generalizations of
Eq.~(\ref{model:marenzoni}) which should remedy this problem.  Of the
models put forward in the literature, we note that Cornwall's proposal
(\ref{model:cornwall}) lies closest to the data.


Fig.~\ref{fig:chisq-ourmodels} shows $\csqdf$ as functions of the
starting point and width (in number of points) of the fits, for fits
to Models A--C.  Of these three, only Model A is able to account
properly for the infrared behavior of the gluon propagator, while
Models B and C yield values for $\csqdf$ of 14 and 12 respectively.
All the models give reasonable fits to the data for intermediate
momentum ranges.

Fit parameters for Model A are illustrated in
Fig.~\ref{fig:params-modelA}.  All the parameters, in particular $M$
and $\alpha$, are well determined and stable over the most interesting
regions (fits with a large number of points, including the infrared).
In the ultraviolet region alone, all the parameter values become
unstable.  This is expected, since we found in
Section~\ref{sec:asymptotics} that a 2-parameter form is sufficient to
describe the data in this region.  There the 4-parameter forms, Models
A--C, will be poorly constrained.

\begin{figure}[p]
\begin{center}
\epsfysize=11.6truecm
\leavevmode
\rotate{\vbox{\epsfbox{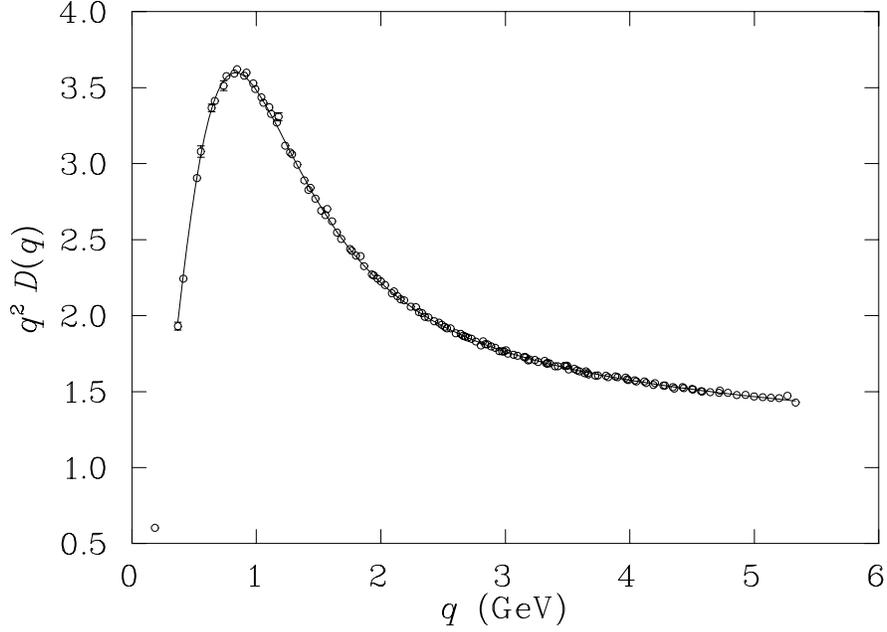}}}
\end{center}
\caption{The gluon propagator multiplied by $q^2$, with nearby
points averaged.  The line illustrates our best fit of Model A
defined in (\protect\ref{modelA}).  The fit is performed over
all points shown, excluding the one at the lowest momentum value, which
may be sensitive to the finite volume of the lattice.
The scale is taken from the value for the string tension quoted in
Ref.~\protect\cite{bs}.
}
\label{fig:fit-modelA}
\end{figure}

\begin{figure}[p]
\begin{center}
\epsfysize=11.6truecm
\leavevmode
\rotate{\vbox{\epsfbox{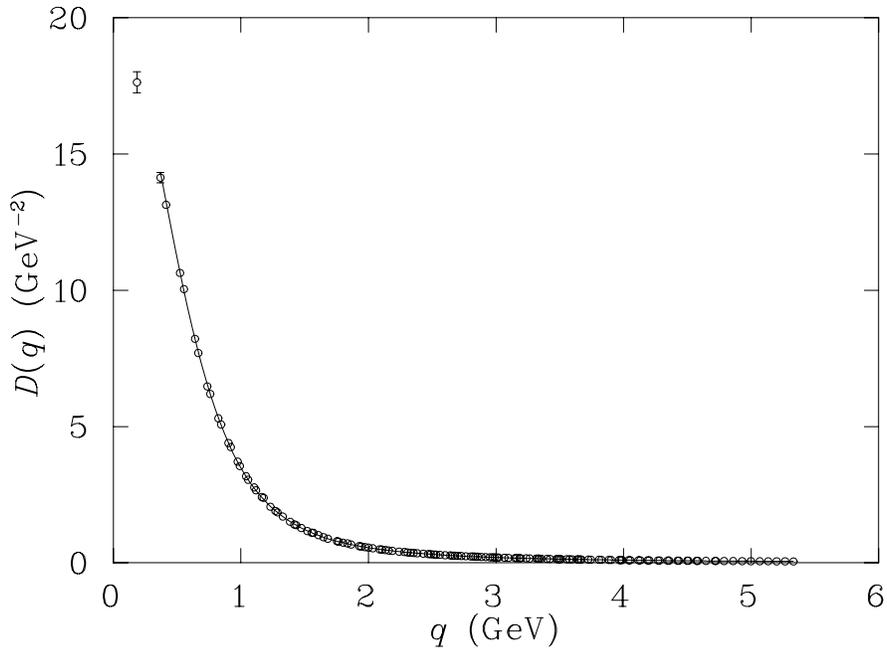}}}
\end{center}
\caption{The gluon propagator in physical units.  The line illustrates
our best fit of Model A, as in Fig.~\protect\ref{fig:fit-modelA}.}
\label{fig:fit-phys-modelA}
\end{figure}

\begin{figure}[p]
\begin{center}
\epsfysize=11.6truecm
\leavevmode
\rotate{\vbox{\epsfbox{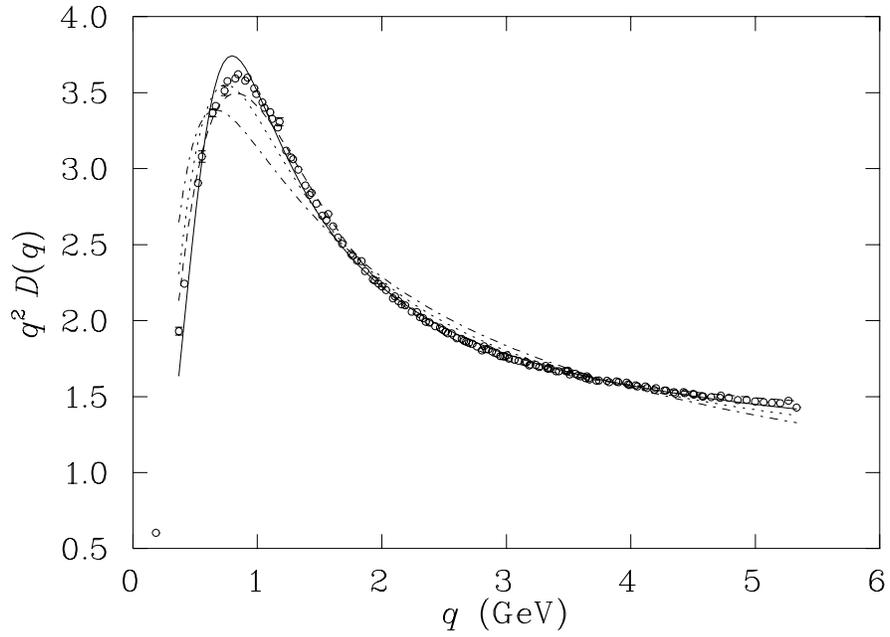}}}
\end{center}
\caption{The gluon propagator multiplied by $q^2$, with nearby points
averaged.  The lines illustrate the best fit of various other models.
The solid line is Model B, the dotted line
is Cornwall I, the dash-dotted line is
Marenzoni and the dashed line is Model C.
}
\label{fig:fit-othermodels}
\end{figure}

Figs.~\ref{fig:fit-modelA} and \ref{fig:fit-phys-modelA} show the best
fit of Model A.  We see that this provides a near perfect fit to the
data.  The optimal fit parameters are shown in Table
\ref{tab:fitparams}.  Fig.~\ref{fig:fit-othermodels} shows fits of
several other models, which we can see fail to account properly for
the data.

Table~\ref{tab:fitparams} also shows the parameter values for
Marenzoni {\em et
al'\/}s model (\ref{model:marenzoni}) and Cornwall's model
(\ref{model:cornwall}).  The values quoted are for fits to all the
available data, while the errors denote the spread in parameter values
resulting from varying the fitting range.  The statistical errors are
in all cases much smaller than the systematic errors associated with
varying the fit regime.  In the case of Model C, the variation in
parameter values becomes unstable in the usual fitting ranges;
in order to avoid this problem, we have chosen a more restricted set
of fitting ranges to evaluate the uncertainties than for the other
models.


In order to determine the dependence of our models on the exact
functional form used to regulate the ultraviolet term in the infrared,
we have performed fits to Eqs.~(\ref{modelA}) and (\ref{modelB})
(Models A and B) with $L(q^2,M)\to L(q^2,\sqrt{2}M)$ and with
$L(q^2,M)\to L(q^2,M/\sqrt{2})$, altering the relationship between
masses in the infrared and ultraviolet terms.  
This turns out to have a significant
effect both on $\csqdf$ and on the values for the fit parameters.  
The quality of the fit deteriorates substantially as
$M\to\sqrt{2}M$ in the logarithm, while it improves slightly as $M\to
M/\sqrt{2}$.  The value for the exponent $\alpha$ changes by more than
$2\sigma$, and this feeds through to the other parameters, although
the value for $M$ remains approximately within $1\sigma$ of its
original value.  The relative performance of Model A and Model B is
not affected, and Model A remains clearly the preferred model of these
two.

\section{Discussion and conclusions}
\label{sec:discuss}

\subsection{Comparison of different models}
\label{sec:discuss-models}

We find that none of the models from the literature give a
satisfactory fit to the data.  It can be argued that Stingl's form
(\ref{model:stingl}) is only supposed to be valid in the deep
infrared.  We do not have sufficient data in this region, or control
over the volume dependence of the data at our lowest momentum values,
to be able to distinguish between the performance of the various
models in this region alone (i.e., the first 10 points in our fits).
All models give a reasonable $\csqdf$ when we fit to only the first
10--20 points.  A
generalisation of Stingl's form has been used to fit to lattice data at
high momentum values \cite{aiso}, but these results are not directly
comparable to ours, since they are obtained for a different choice of
gauge and do not include the infrared region ($\qhat_{\rm min} = 1$
GeV).  In the case of Cornwall's proposal (\ref{model:cornwall}) it
should also be mentioned that this form was derived using a
gauge-invariant `pinch technique', and may not be directly comparable
to our Landau gauge results.

We have found that the data can be adequately described by two terms:
one governing the ultraviolet behavior according to the one-loop
perturbative formula, and the other providing the infrared behavior.
The infrared term is proportional to $(q^2+M^2)^{-\alpha}$ with
$\alpha\approx 3$.  It should be emphasized at this point that the
performance of Model A, and in particular the value of the exponent
$\alpha$, depends substantially on the exact form chosen for the
logarithmic function $L(q^2,M)$ in order to regulate the ultraviolet
term in the infrared.  Given a particular form for $\Duv$, all the
parameter values are very stable and lend credence to Model A as
correctly encapsulating the lattice results.  It should be noted,
however, that although the 1-loop perturbative form does provide an
adequate fit in the ultraviolet, we found in Sec.~\ref{sec:asym-fits}
that the parameter values are not stable.  The 2-loop form should be
an improvement on this.  The third model we considered, using an
exponential function rather than a `mass' term to describe the
infrared behavior, was clearly unsatisfactory.

Our approach here differs significantly from those of previous studies
\cite{bps,mms}.  Firstly, in order to reduce the effect of
lattice artefacts at high momenta, we use the momentum variable $q$
defined in Eq.~(\ref{eq:latt-momenta}) rather than the `naive' momentum
variable $\qhat$.  We believe this approach has been justified by the
verification of scaling in
Section~\ref{sec:asymptotics}.  A similar approach has been used in a
recent study of the three-gluon vertex \cite{ggg-orsay}.  Furthermore,
we select an improved and larger set of momenta to the ones used in
those previous 
studies.  With this in mind, it should nevertheless be possible to
make at least an approximate comparison between the results at small to
intermediate momenta.

Both previous studies \cite{bps,mms} fit their data to Marenzoni's form
(\ref{model:marenzoni}) or special cases of this model (with $M=0$ or
$\alpha=0$), which we have found does not account satisfactorily for
the data.  In addition, in Ref.~\cite{bps} the low-momentum data are fitted
to the Gribov form (\ref{model:gribov}).  The latter form fails to
provide us with any fit which would make a comparison of parameter
values meaningful.  However, we may compare the values we obtain for
the parameter $\alpha$ in the Marenzoni form of 
Eq.~(\ref{model:marenzoni}) with those of
Refs.~\cite{bps,mms}.  Although $\csqdf$ for fits to all the data
with this model is very high, the value for $\alpha$ is reasonably stable
over a large region, including fits where $\csqdf \sim 1$.  We
find $\alpha \sim 0.3$, in agreement with the value quoted in
Ref.~\cite{bps}.  This is inconsistent with the value of $\sim 0.5$ quoted
in Ref.~\cite{mms}.  However, this value is obtained by fitting only to data
in the infrared region.  If we restrict ourselves to the same region,
we also obtain a value of $\alpha \sim 0.5$.  Hence, when repeating
the analysis of Refs.~\cite{bps,mms}, we find results consistent with
theirs. 

We find that Model A provides a fit to the data throughout the entire
available momentum range.  However, we are unaware of any current
physical interpretation of this model, in contrast to the models
arising from the approximate analytical studies by Gribov, Stingl and
Cornwall \cite{gribov,stingl,cornwall}.

\subsection{Finite volume effects}
\label{sec:finite-vol}

The asymmetry of the lattices, with $L_t=3L_s$ for the small
lattice and $L_t=2Ls$ for the large and fine lattices, is one of the
measures used to
assess finite volume effects.  By comparing momenta along
the time axis with momenta along the spatial axes, we find that finite
volume errors are small on the large lattice, even at low momentum
values.  A `cone' cut along the diagonal in momentum space is imposed
for the smaller lattices to remove finite volume effects, but this cut
is not found to be necessary for the large lattice.

Inspection of the tensor structure reveals some residual finite
volume effects in the order of 10--15\% at the lowest momentum
values. 
Apart
from these 4--6 points, finite volume effects are negligible.
Excluding these points does not change the parameter values.  Nor will
the relative performance of our models be affected.
Implementing the `cone' cut on the large lattice
will have a similar effect to excluding these points.

Comparing the data at low momenta for the two lattices at $\beta=6.0$,
we find that the value of the gluon propagator decreases with
increasing volume.  This opens up the possibility that in the
infinite-volume limit, the propagator may be strongly suppressed or
even vanishing at extremely small momenta, as suggested by Gribov and
Stingl \cite{gribov,stingl}.  Recent studies at strong coupling and in
lower dimensions \cite{furui,cucchieri-ir,cucchieri-3d} lend some
support to this possibility.

\subsection{Finite lattice spacing effects}
\label{sec:finite-a}

The kinematic correction $\qhat\to q$ gives a large reduction in
finite lattice spacing anisotropy at high momentum values, but does
not remove this anisotropy completely.  A `cylinder' cut along the
diagonal in momentum space is imposed on all lattices to remove this
residual anisotropy.

We have verified scaling of the gluon propagator for momenta $q > 1.3$
GeV between $\beta=6.0$ and $\beta=6.2$.  This scaling is dependent on
the kinematic correction $\qhat\to q$.  If $\qhat$ is used as the
momentum variable, scaling fails, even after the `cylinder' and `cone'
cuts are imposed.  We are currently working on using improved actions
\cite{improve,landau-imp} to reduce or remove finite lattice spacing
effects.

\subsection{Conclusion}
\label{sec:conclude}

We have calculated the gluon propagator on a large volume lattice and
verified that finite volume effects are under control.  Finite volume
effects in the order of 10\% are found for the very lowest momentum
values, but become insignificant for $q > 600$ MeV.  Finite lattice
spacing effects are handled by using the kinematic correction
$\qhat_\mu \to q_\mu = (2/a)\sin(\qhat_\mu a/2)$, and by selecting
momenta along the 4-dimensional diagonal.  Scaling is verified
between $\beta=6.0$ and 6.2.

The propagator is found to be well represented by the functional form
$D(q^2) = \Dir+\Duv$, where $\Dir = AM^{2\alpha}(q^2+M^2)^{-(1+\alpha)}$ and $\Duv$ is an
infrared regulated version (see Eqs.~(\ref{eq:uvlog-def}) and
(\ref{modelA})) of the one-loop asymptotic form defined in
Eq.~(\ref{model:asymptotic}).  Our best estimate for the parameter
$\alpha$ is $\alpha=2.2\err{0.1}{0.2}\err{0.2}{0.3}$, where the second set of
errors represents the systematic uncertainty arising from the choice
of infrared regulator for $\Duv$.
Using the regulator given in Eq.~(\ref{eq:uvlog-def}), our best estimates for the parameters
$M$ and $A$ are $M=(1020\pm 100\pm 25)$ MeV and $A=9.8\err{0.1}{0.9}$,
where the second set of errors in $M$ represents the statistical
uncertainty in the lattice spacing quoted in Table~\ref{tab:sim-params}.

Among the issues still under consideration are an extrapolation of
$D(q^2)$ to infinite volume at low $q^2$, as well as an evaluation of
the effect of Gribov copies and of the gauge dependence of the gluon
propagator.  Work is also in progress to calculate the gluon
propagator using improved actions, thereby reducing
finite lattice spacing effects and allowing simulations on larger
physical volumes.

\section*{Acknowledgments} 

Financial support from the Australian Research Council is gratefully
acknowledged.  The numerical work was mainly performed on a Cray T3D
based at EPCC, University of Edinburgh, using UKQCD Collaboration CPU
time under PPARC Grant GR/K41663. Also, CP acknowledges further
support from PPARC through an Advanced Fellowship.  We thank G.~Cvetic
for pointing out an error in the gluon anomalous dimension in our
original manuscript.

\begin{figure}[p] 
\begin{center}
\setlength{\unitlength}{1cm}
\setlength{\fboxsep}{0cm}
\begin{picture}(14,18)
\end{picture}
\end{center}
\vspace{-4cm}
\caption{$\chi^2$ per degree of freedom for fits of Model A (top left),
Model B (top right) and Model C (bottom).  The ``Fit start'' axis
indicates the starting point for the fit, while the ``Fit width'' axis
indicates the number of points included in the fit.  The most
inclusive fits are in the near right-hand corner, with the smallest
value for the starting point and the largest number of points
included.  We see that Model A is stable over a wide variety
of fitting ranges, while the other two models fail to account properly
for the data in the infrared.}
\label{fig:chisq-ourmodels}
\end{figure}

\begin{figure}[p] 
\begin{center}
\setlength{\unitlength}{1.0cm}
\setlength{\fboxsep}{0cm}
\begin{picture}(14,18)
\end{picture}
\end{center}
\vspace{-3cm}
\caption{Stability plots for Model A.
All the parameter values are stable over the region of interest.}
\label{fig:params-modelA}
\end{figure}

\begin{table}[p]
\begin{center}
\leavevmode
\begin{tabular}{lddcccccl}
Name &$\beta$ & $a\sqrt{K}$  &$a^{-1}$ (GeV) &Volume &$N_{\rm config}$ & Separation &
 $\theta_{max}$ & $\bra U\ket$ \\
\hline
Small    &6.0 & 0.2265(55) &1.885(45)   &$16^3\times 48$ &125 & 800 & $10^{-12}$ & 0.860939(31) \\
Large    &6.0 & 0.2265(55) &1.885(45)  &$32^3\times 64$ &75 & 1000 & $10^{-12}$ & 0.861793(15) \\
Fine     &6.2 & 0.1619(19) &2.637(30)  &$24^3\times 48$ &223 & 2400 & $10^{-12}$ & 0.873948(15)\\
\end{tabular}
\end{center}
\caption{Simulation parameters.  The values for the string tension
$a\protect\sqrt{K}$ are taken from Ref.~\protect\cite{bs}, and the
lattice spacings are calculated using the `physical' value
$\protect\sqrt{\protect\sigma}=427$ MeV for the string tension.  The
separation is the total number of updates (Cabibbo--Marinari or
over-relaxation) separating the configurations.  $\protect\bra
U\protect\ket$ is the average link $\protect\sum_{x,\mu}\Re\Tr
U_{\mu}(x)/(4VN_c)$}
\label{tab:sim-params}
\end{table}

\begin{table}[p]
\begin{tabular}{cc|d|d|d|d} 

 & & \multicolumn{2}{c|}{Theoretical prediction} &
 \multicolumn{2}{c}{This simulation} \\
{\bf $[\qhat_x,\qhat_y,\qhat_z,\qhat_t]$} & Components
 & Using $\qhat$ & Using $q$ & Using $A$ &  Using $A'$ \\ \hline

[2,1,0,0]
& (1,1)/(1,2) &  -0.5 &  -0.509796 &  -0.509796 &  -0.519783 \\
& (1,1)/(2,2) &  0.25 &   0.259892 &   0.259892 & 0.259892  \\
& (1,1)/(3,3) &   0.2 &   0.206281 &   0.204(8) & 0.204(8)  \\
& (1,1)/(4,4) &   0.2 &   0.206281 &   0.199(9) & 0.199(9)  \\
& (1,2)/(2,2) &  -0.5 &  -0.509796 &  -0.509796 & -0.5  \\ 
& (1,2)/(3,3) &  -0.4 &  -0.404634 &   -0.40(2) & -0.38(2) \\ \hline

[4,1,0,0]
& (1,1)/(1,2) & -0.25 &  -0.275899 &  -0.275899  & -0.331821 \\
& (1,1)/(2,2) &  0.0625 &  0.0761205 & 0.0761205 &  \\
& (1,2)/(2,2) & -0.25 &  -0.275899 &  -0.275899  & -0.229402 \\
& (1,2)/(3,3) & -0.2353 & -0.256383 & -0.277(12) & -0.231(10) \\ \hline

[4,2,0,0]
& (1,1)/(1,2) &  -0.5 &  -0.541196 &  -0.541196 & -0.585786 \\
& (1,1)/(2,2) &  0.25 &   0.292893 &   0.292893 &  \\
& (1,1)/(3,3) &   0.2 &   0.226541 &    0.22(1) &  \\
& (1,2)/(2,2) &  -0.5 &  -0.541196 &  -0.541196 &      -0.5 \\ \hline

[2,1,1,0]
& (1,1)/(1,2) &    -1 &   -1.01959 &    -1.01(2) &   -1.03(2) \\
& (1,1)/(2,2) &   0.4 &   0.412562 &   0.411(15) &  \\
& (1,1)/(3,3) &   0.4 &   0.412562 &   0.418(14) &  \\
& (1,2)/(2,2) &  -0.4 &  -0.404634 &  -0.407(10) & -0.398(11) \\ \hline

[4,2,1,0]
& (1,1)/(1,2) & -0.625 &   -0.681848 &  -0.678(9) & -0.743(10) \\
& (1,1)/(2,2) & 0.2941 &    0.342911 &   0.339(7) &  \\
& (1,1)/(2,3) &   -2.5 &    -2.47137 &    -2.3(4) &    -2.5(5) \\
& (1,3)/(3,3) &   -0.2 &   -0.213397 & -0.208(10) & -0.187(11) \\ \hline

[4,2,1,1/3]
& (1,1)/(1,2) &  -0.6389 &  -0.697656 &  -0.695(9)  &  -0.750(10)  \\
& (1,1)/(2,2) &   0.2987 &   0.348094 &   0.348(8)  &              \\
& (1,1)/(4,4) &   0.2434 &   0.275796 &  0.288(13)  &              \\
& (1,2)/(2,2) &  -0.4675 &  -0.498947 &  -0.500(7)  &   -0.464(7)  \\
& (1,3)/(2,2) &  -0.2338 &  -0.254361 &   -0.25(2)  &    -0.20(2)  

\end{tabular}
\caption{Tensor structure for the small lattice. $\qhat$ is in units
of $2\pi/L_s$, where $L_s$ is the spatial length of the lattice.  The
theoretical predictions are the values for the ratios one obtains from
(\protect\ref{eq:landau-prop}), and from
(\protect\ref{eq:landau-prop}) with $q\to\qhat$.  The numbers in
brackets are the statistical uncertainties in the last digit(s).
Where no error is quoted, the statistical uncertainty is less than
$10^{-6}$.  The values obtained using the asymmetric gluon field
definition $A'$ are only shown where they differ from the value using
$A$.}
\label{tab:tensor-small}
\end{table}

\begin{table}[p]
\begin{tabular}{lc|d|d|d}

 & & \multicolumn{2}{c|}{Theoretical prediction} &
This simulation \\
{\bf $[\qhat_x,\qhat_y,\qhat_z,\qhat_t]$} & Components
 & Using $\qhat$ & Using $q$ & Using $A$ \\ \hline

[2,1,0,0]
& (1,1)/(1,2) &     -0.5 &     -0.502419 &     -0.502419  \\
& (1,1)/(2,2) &     0.25 &      0.252425 &      0.252425  \\
& (1,1)/(3,3) &      0.2 &      0.201549 &      0.217(13)  \\
& (1,2)/(3,3) &     -0.4 &     -0.401157 &     -0.43(3)  \\ \hline

[8,4,0,0]
& (1,1)/(1,2) &     -0.5 &     -0.541196 &     -0.541196  \\
& (1,1)/(2,2) &     0.25 &      0.292893 &      0.292893  \\
& (1,1)/(3,3) &      0.2 &      0.226541 &      0.21(1)  \\ \hline

[8,4,2,0]
& (1,1)/(1,2) &   -0.625 &     -0.681848 &     -0.691(12)  \\
& (1,1)/(2,2) &   0.2941 &      0.342911 &      0.351(11)  \\
& (1,2)/(2,2) &  -0.4706 &     -0.502914 &      -0.508(9)  \\
& (1,3)/(3,3) &     -0.2 &     -0.213397 &    -0.223(15) \\ \hline

[8,2,1,1/2]
& (1,1)/(1,2) &  -0.3281 &     -0.362996 &     -0.360(6)  \\
& (1,1)/(4,4) &  0.07609 &     0.0914336 &      0.085(4)  \\
& (1,2)/(2,2) &  -0.2452 &     -0.269425 &     -0.265(5)  \\
& (1,3)/(2,2) &  -0.1226 &     -0.135364 &    -0.134(11)

\end{tabular}
\caption{Tensor structure for the large lattice.  $\qhat$ is given in
units of $2\pi/L_s$, where $L_s$ is the spatial length of the lattice.
Since the spatial length of this lattice is twice that of the small
lattice, the values for $\qhat$ must be multiplied by 2 when comparing
these values with those of Table~\protect\ref{tab:tensor-small}.}
\label{tab:tensor-large}
\end{table}

\begin{table}[p]
\begin{tabular}{lc|d|d|d|d}

 & & \multicolumn{2}{c|}{Theoretical prediction} &
 \multicolumn{2}{c}{This simulation} \\
{\bf $[\qhat_x,\qhat_y,\qhat_z,\qhat_t]$} & Components
 & Using $\qhat$ & Using $q$ & Using $A$ &  Using $A'$ \\ \hline

[2,1,0,0]
& (1,1)/(1,2) &    -0.5 &    -0.504314 &    -0.504315  & -0.508666 \\
& (1,1)/(2,2) &    0.25 &     0.254333 &     0.254333  & \\
& (1,2)/(2,2) &    -0.5 &    -0.504314 &    -0.504315  & -0.5 \\
& (1,2)/(3,3) &    -0.4 &    -0.402058 &    -0.405(14)  & -0.402(14) \\ \hline

[6,1,0,0]
& (1,1)/(1,2) & -0.1667 &    -0.184592 &    -0.184592  & -0.232673 \\
& (1,2)/(2,2) & -0.1667 &    -0.184592 &    -0.184592  & -0.146447 \\
& (1,2)/(3,3) & -0.1622 &    -0.178509 &    -0.183(6)  & -0.145(5) \\ \hline

[6,3,0,0]
& (1,1)/(1,2) &   -0.5 &   -0.541196 &   -0.541196 &  -0.585786 \\
& (1,1)/(2,2) &   0.25 &    0.292893 &    0.292893 &  \\
&(1,1)/(3,3) &    0.2 &    0.226541 &    0.226(7) &  \\
& (1,2)/(2,2) &   -0.5 &   -0.541196 &   -0.541196 &       -0.5 \\
\hline

[6,3,1,0]
& (1,1)/(1,2) & -0.5556 &    -0.604157 &     -0.605(4)  & -0.655(4) \\
& (1,1)/(2,2) &  0.2703 &     0.316193 &     0.318(4)  &  \\
& (1,3)/(3,3) & -0.1333 &    -0.142774 &     -0.149(8)  & -0.118(8) \\ \hline

[6,3,1,1]
& (1,1)/(1,2) & -0.6111 &    -0.667118 &    -0.666(7)  & -0.717(8) \\
& (1,1)/(4,4) &  0.2391 &      0.27208 &      0.270(9)  &  \\
& (1,2)/(2,2) & -0.4737 &    -0.506668 &    -0.500(5)  & -0.464(5) \\
& (1,3)/(2,2) & -0.1579 &    -0.172815 &    -0.188(14)  & -0.147(12) \

\end{tabular}
\caption{Tensor structure for the fine lattice}
\label{tab:tensor-fine}
\end{table}

\begin{table}[p]
\begin{tabular}{cc|d|d}
$[\qhat_x,\qhat_y,\qhat_z,\qhat_t]$ & Components
 & Ratio according to (\ref{eq:landau-prop}) & This simulation \\ \hline

[1,0,0,0]
& (2,2)/(3,3) &     1 &    1.01(5)  \\       
& (2,2)/(4,4) &     1 &    1.24(6)  \\       
& (3,3)/(4,4) &     1 &    1.23(6)  \\ \hline

[0,0,1,0]
& (1,1)/(4,4) &     1 &    1.25(5)  \\       
& (2,2)/(4,4) &     1 &    1.32(6)  \\ \hline

[1,0,0,1/3]
& (1,1)/(2,2) &   0.101034 &  0.083(4)  \\
& (2,2)/(4,4) &    1.11239 &    1.35(7)  \\
& (3,3)/(4,4) &    1.11239 &    1.36(6)  \\ \hline

[1,0,0,2/3]
& (1,1)/(2,2) &   0.309218 &    0.275(13)  \\
& (2,2)/(4,4) &    1.44763 &    1.63(7)  \\
& (3,3)/(4,4) &    1.44763 &    1.61(7)  \\ \hline

[1,1,0,0]
& (3,3)/(4,4) &     1 &    1.10(5)  \\ 

[1,0,1,0]
& (2,2)/(4,4) &     1 &    1.17(6)  \\

[0,1,1,0]
& (1,1)/(4,4) &     1 &    1.05(5)  \\ \hline

[1,0,0,1]
& (1,1)/(4,4) &     1 &      1 \\
& (2,2)/(1,1) &     2 &    1.98(10)  \\
& (3,3)/(4,4) &     2 &    2.22(10)  \\ \hline

[0,0,1,1]
& (1,1)/(3,3) &     2 &    2.25(10)  \\
& (1,1)/(4,4) &     2 &    2.25(10)  \\
& (2,2)/(4,4) &     2 &    2.13(10) 

\end{tabular}
\caption{Tensor structure for low momentum values on the small
lattice.}
\label{tab:tensor-small-lowq}
\end{table}

\begin{table}[p]
\begin{tabular}{cc|d|d}
$[\qhat_x,\qhat_y,\qhat_z,\qhat_t]$ & Components
 & Ratio according to (\ref{eq:landau-prop}) & This simulation \\ \hline

[1,0,0,0]
& (2,2)/(3,3) &     1 &   0.95(3)  \\
& (2,2)/(4,4) &     1 &    1.20(4)  \\
& (3,3)/(4,4) &     1 &     1.26(4)  \\ \hline

[2,0,0,0]
& (2,2)/(3,3) &     1 &    1.00(4)  \\
& (2,2)/(4,4) &     1 &    1.08(4)  \\
& (3,3)/(4,4) &     1 &    1.07(4)  \\ \hline

[1,0,0,1/2]
& (1,1)/(2,2) &   0.200687 &     0.170(6)  \\
& (1,1)/(3,3) &   0.200687 &   0.187(6)  \\
& (2,2)/(4,4) &    1.25107 &    1.47(5)  \\
& (3,3)/(4,4) &    1.25107 &    1.34(4) 

\end{tabular}
\caption{Tensor structure for low momentum values on the fine
lattice.}
\label{tab:tensor-fine-lowq}
\end{table}

\begin{table}[p]
\begin{tabular}{lc|d|d}

$[\qhat_x,\qhat_y,\qhat_z,\qhat_t]$ & Components
 & Ratio according to (\ref{eq:landau-prop}) & This simulation \\ \hline

[1,0,0,0]
& (2,2)/(3,3) &     1 &   0.97(5)  \\
& (2,2)/(4,4) &     1 &    1.13(6)  \\
& (3,3)/(4,4) &     1 &    1.17(6)  \\ 

[0,1,0,0]
& (1,1)/(4,4) &    1 &   1.13(7)  \\
& (3,3)/(4,4) &    1 &   1.17(8)  \\ 

[0,0,1,0]
& (1,1)/(4,4) &    1 &   1.05(6)  \\
& (2,2)/(4,4) &    1 &   1.08(6)  \\ \hline

[2,0,0,0]
& (2,2)/(3,3) &     1 &    0.97(7)  \\
& (2,2)/(4,4) &     1 &    1.10(7)  \\
& (3,3)/(4,4) &     1 &    1.14(6)  \\ 

[0,2,0,0]
& (1,1)/(4,4) &    1 &   1.09(6)  \\
& (3,3)/(4,4) &    1 &   1.02(7)  \\ \hline

[1,0,0,1/2]
& (1,1)/(2,2) &   0.200386 &   0.182(10)  \\
& (1,4)/(2,2) &  -0.400289 &  -0.36(2)  \\
& (2,2)/(4,4) &     1.2506 &    1.37(7)  \\
& (3,3)/(4,4) &     1.2506 &    1.34(7)  \\ \hline

[0,0,1,1/2]
& (3,3)/(1,1) &   0.200386 &   0.196(10) \\
& (3,4)/(1,1) &  -0.400289 &  0.39(2) \\
& (1,1)/(4,4) &    1.2506 &   1.27(7) \\
& (2,2)/(4,4) &    1.2506 &   1.33(8)

\end{tabular}
\caption{Tensor structure for the large lattice, low
values of $\qhat$.  Note that $\qhat$ is given in units of $2\pi/L_s$,
so that eg.\ $\qhat=[2,0,0,0]$ corresponds to $\qhat=[1,0,0,0]$ for
the small lattice.}
\label{tab:tensor-large-lowq}
\end{table}

\begin{table}[p]

\begin{tabular}{c|d|d|d}

Components & Small lattice & Large lattice & Fine lattice \\ \hline
(1,1)/(2,2) &   0.93(6)  &   1.05(8)  &   0.98(4)   \\
(1,1)/(3,3) &   1.03(6)  &   0.94(8)  &   0.98(4)   \\
(2,2)/(3,3) &   1.10(7)  &   0.90(8)  &    1.00(5)   \\
(1,1)/(4,4) &   3.09(19) &   1.42(11) &     2.06(9)   \\
(2,2)/(4,4) &   3.31(20) &   1.35(12) &    2.10(10)   \\
(3,3)/(4,4) &   3.00(18) &   1.51(11) &    2.09(9) 

\end{tabular}
\caption{Ratios of the diagonal components of $D_{\mu\nu}(q=0)$ for
all three lattices.}
\label{tab:tensor-zeromom}
\end{table}

\begin{table}[p]

\begin{tabular}{c|cc|cc|c|r|r|rc|}

Lattice & \multicolumn{2}{c}{$q_{min}\,,\,q_{max}$ ($a^{-1}$)} & 
\multicolumn{2}{c}{$q_{min}\,,\,q_{max}$ (GeV)} & No.\ of points & 
$\csqdf$ & $Z$ & $\Lambda a$ & $\Lambda$ (GeV) \\ \hline

$\beta=6.0$
 & 1.47 & 2.78 & 2.72 & 5.14 & 27 & 1.48 & 2.140 & 0.399 & 0.752 \\
$16^3\times 48$
 & 1.59 & 2.78 & 2.94 & 5.14 & 25 & 1.43 & 2.162 & 0.387 & 0.730 \\
 & 2.12 & 2.78 & 3.92 & 5.14 & 15 & 1.13 & 2.151 & 0.394 & 0.743 \\ \hline
$\beta=6.0$
 & 1.53 & 2.83 & 2.83 & 5.23 & 69 & 1.42 & 2.159 & 0.387 & 0.729 \\
$32^3\times 64$
 & 1.53 & 2.76 & 2.83 & 5.10 & 67 & 1.34 & 2.157 & 0.387 & 0.730 \\
 & 2.10 & 2.76 & 3.89 & 5.10 & 29 & 1.28 & 2.184 & 0.373 & 0.703 \\
 & 2.12 & 2.76 & 3.91 & 5.10 & 27 & 1.10 & 2.220 & 0.354 & 0.667 \\
 & 2.12 & 2.83 & 3.91 & 5.23 & 29 & 1.29 & 2.223 & 0.350 & 0.660 \\ \hline
$\beta=6.2$
 & 1.09 & 2.83 & 2.83 & 7.36 & 53 & 1.33 & 2.286 & 0.275 & 0.726 \\
$24^3\times 48$
 & 1.09 & 2.44 & 2.83 & 6.34 & 43 & 1.30 & 2.274 & 0.279 & 0.736 \\
 & 1.09 & 2.00 & 2.83 & 5.20 & 29 & 0.81 & 2.222 & 0.297 & 0.782 \\
 & 1.49 & 2.83 & 3.87 & 7.36 & 41 & 1.02 & 2.341 & 0.253 & 0.668 \\
 & 1.49 & 2.00 & 3.87 & 5.20 & 17 & 0.78 & 2.212 & 0.301 & 0.793

\end{tabular}

\caption{Parameter values and $\chi^2$ for fits to the asymptotic form
(\protect\ref{model:asymptotic}).  Note that in this table,
$Z$ is actually $Z_3(\mu,a)Z$ of Eq.~(\protect\ref{model:asymptotic}).
The fits are to data surviving
the cylindrical and cone cuts, except for the large lattice, where
only the cylindrical cut has been applied.}
\label{tab:asymptotic_fits}
\end{table}

\begin{table}[p]

\begin{tabular}{c|c|cr@{--}lc|cr@{--}lc}

Model & $\chi^2_{\rm all}/N_{df}$ & $\chi^2_{\rm min}/N_{df}$ &
\multicolumn{2}{c}{Range ($qa$)} & $N_{\rm fit}$ & $\chi^2_{\rm IR}/N_{df}$ &
\multicolumn{2}{c}{Range ($qa$)} & $N_{\rm fit}$ \\ \hline
Gribov & 827 & 0.31 & 0.28 & 0.39 & 5 \\
Stingl & 838 & 0.44 & 0.28 & 0.45 & 8 & 1.03 & 0.28 & 0.49 & 10 \\
Marenzoni & 163 & 0.79 & 1.11 & 1.47 & 24 
 & 1.20 & 0.20 & 0.59 & 18 \\
Cornwall I & 50 & 0.67 & 0.99 & 1.47 & 30
 & 1.01 & 0.20 & 0.62 & 20 \\
Cornwall II & 89 & 0.49 & 1.26 & 1.47 & 14 &
 0.69 & 0.28 & 0.45 & 8 \\
Cornwall III & 38 & 0.64 & 1.13 & 1.33 & 23 & 1.05 & 0.20 & 0.48 & 11
\end{tabular}
\caption{$\chi^2$ per degree of freedom for fits to the models
(\protect\ref{model:lita})--(\protect\ref{model:litz}).
$\chi^2_{\rm all}$ is the $\chi^2$ for the maximum available
fitting range.  $\chi^2_{\rm min}$ corresponds to the minimum value obtained for
$\csqdf$, ``Range'' is the corresponding fitting range, and
$N_{\rm fit}$ is the number of points included in that range.
$\chi^2_{\rm IR}$ refers to the widest fitting range starting
in the deep infrared (point 2 or 4) where $\csqdf \protect\lesssim 1$.}
\label{tab:chisq_litmodels}
\end{table}

\begin{table}[p]
\begin{tabular}{c|d|dddd}
Model & $\csqdf$ & $Z$ & $A$ & $M$ & $\alpha$ or $\Lambda$
\\ \hline
Marenzoni & 163 & 2.41\err{0}{12} & & 0.14\err{4}{14}
 & 0.29\err{6}{2} \\
Cornwall I & 50.3 & 6.5\err{7}{9} & & 0.24\err{3}{16} &
0.27\err{7}{7}  \\
Model C & 13.9 & 2.55\err{147}{0} & 1.07\err{8}{97} & 0.23\err{1}{22}
 & 0.53\err{4}{29} \\
Model A & 1.40 & 2.01\err{4}{5} & 9.84\err{10}{86} & 0.54\err{5}{5}
 & 2.17\err{11}{19} \\
Model $\rm A_2$ & 2.16 & 2.03\err{2}{1} & 8.85\err{44}{0} 
 & 0.51\err{1}{2} \\
Model B & 12.1 & 2.09\err{0}{21} & 2.29\err{17}{39} & 0.38\err{23}{1}
 & 1.09\err{46}{7}
\end{tabular}
\caption{Parameter values in lattice units for fits of models
(\protect\ref{model-first})--(\protect\ref{model-last}).  The
values quoted are for fits to the entire set of data.  The errors
denote the uncertainty in the last digit(s) of the parameter values
which results from varying the fitting range.  The fitting ranges
considered when evaluating the uncertainties are those with a minimum
of 40 points included and with the minimum value for $qa$ no larger
than 0.99 (point number 40), corresponding to $q_{\rm min}\leq 1.86$
GeV.  For Model C, the fitting ranges have been restricted to minimum
values for $qa$ no larger than 0.62 (point number 20), in order to
obtain meaningful uncertainties.  Model $\rm A_2$ denotes Model A with
$\alpha$ fixed to 2.
Recall that the inverse lattice spacing for this lattice is 1.885 GeV.}
\label{tab:fitparams}
\end{table}

\end{document}